\documentclass[12pt]{article}

\usepackage{amsmath, amsthm, amsfonts}
\usepackage{graphicx}
\usepackage{amssymb}
\usepackage{mathrsfs}
\usepackage{tikz} 
\usepackage{hyperref,cite}
\hypersetup{colorlinks=true,linkcolor= blue,citecolor=blue}
\usepackage[usenames,dvipsnames]{pstricks}

\usepackage{parskip}
\numberwithin{equation}{section}

\usetikzlibrary{arrows,snakes}
\newcommand{\be}{\begin{equation}}                                                                                          
\newcommand{\ee}{\end{equation}}                                               
\newcommand{\p}{\partial}                                           
\long\def\begincomment#1\endcomment{} 
\definecolor{lightblue}{rgb}{.1,.4,.5}
\definecolor{brown1}{rgb}{.64,.43,.138}

\title{Holographic Brownian Motion in 1+1 Dimensions}
\author{Pinaki Banerjee{\footnote {pinakib@imsc.res.in} } and B. Sathiapalan{\footnote{bala@imsc.res.in}} \\ \\
  \small Institute of Mathematical Sciences\\  
  \small CIT Campus, Taramani, Chennai-600113\\
  \small India }
\date{}

\begin{document}

\hspace{15cm} IMSc/2013/08/6

{\let\newpage\relax\maketitle}

\renewcommand{\baselinestretch}{1.15}\normalsize

\abstract{We study the motion of a stochastic string in the background of a BTZ black hole. In the 1+1 dimensional boundary theory this corresponds to a very heavy external particle (e.g, a quark), interacting with the fields of a CFT at finite temperature, and describing Brownian motion.  The equations of motion for a string in the BTZ background can be solved exactly. Thus we can use holographic techniques to obtain the Schwinger-Keldysh Green function for the boundary theory for the force acting on the quark. We write down the generalized Langevin equation describing the motion of the external particle and calculate the drag and the thermal mass shift. Interestingly we obtain dissipation even at zero temperature for this 1+1 system. Even so, this does not violate boost (Lorentz) invariance because the drag force on a {\em constant} velocity quark continues to be zero.  Furthermore since the Green function is exact, it is possible to write down an effective membrane action, and thus a Langevin equation, located at  a ``stretched horizon''  at an arbitrary finite distance from the horizon.}

\newpage 

\tableofcontents

\section{Introduction}
\label{sec:intro}

AdS/CFT correspondence \cite{Maldacena,GKP,Witten1} has been used quite successfully to study thermal properties such as the viscosity of $ \mathcal{N}=4$ super Yang-Mills theory at finite temperature.  Dissipation and thermal fluctuation are two sides of the same coin as embodied in the famous fluctuation dissipation (FD) theorem. The study of fluctuations using holographic techniques has been done in several papers \cite{Mukund,Son-Teaney,Tong,Pedraza1,Caron-Huot,Ayan,Pedraza2,Pedraza3} and the fluctuation dissipation theorem has been shown to hold. Different techniques \cite{Mukund,Son-Teaney} have been used to address this issue. A very versatile technique
is in terms of Green functions. Son and Teaney \cite{Son-Teaney} have used holographic techniques to calculate Green functions to address these 
questions in the context of Brownian motion of a particle such as a quark.  
\par The fluctuation-dissipation theorem  in the context of Brownian motion has been studied by Kubo \cite{Kubo1,Kubo2} and Mori \cite{Mori1,Mori2} amongst others. Brownian motion can be described as a stochastic process \cite{vanKamp}. In some approximation it is Markovian. If we can assume that velocities at two instants are not correlated, then it is a Markovian process when described in terms of position. Thus one can define a probability $P\left(x(t),t;x(t_0),t_0\right)$ as the conditional probability for the particle to be in position $x(t)$ at time $t$ given that it was at $x(t_0)$ at time $t_0$. One can also write a Fokker Plank equation for $P(x(t),t;x(t_0),t_0)$. On the other hand if we want a finer description one can use the velocity as the variable defining the Markovian process in terms of $P(v(t),t;v(t_0),t_0)$. This is a good approximation as long as the duration of a collision is very small, which is equivalent to saying that acceleration at different instants is uncorrelated. The Fokker-Planck equation in the velocity description is
\be    \label{FP}
{\partial P(v,t)\over \p t} = - {\p \over \p v}a_1(v)  P + {a_2\over 2} {\p ^2 P \over \p v^2}
\ee
Here $a_1={ \langle \Delta v \rangle \over \Delta t}$ and $a_2 ={ \langle (\Delta v)^2\rangle \over \Delta t}$. Here $\Delta v$ is the change
in velocity in time $\Delta t$. 

\par One can obtain these from the related Langevin equation:
\be	\label{Lang}
m \dot v = -\gamma v + \xi (t)
\ee
where $\xi(t)$ is the random force that is responsible for the fluctuations, obeying $\langle \xi (t) \xi (t')\rangle = \Gamma \delta (t-t')$ and $\langle \xi (t)\rangle =0$. $v(t_0)=v_0$ is the initial condition. Thus $a_1 =\langle v(\Delta t) - v_0\rangle = -{\gamma \over m}v_0 \Delta t$.
From the solution of the Langevin equation (taking $t_0=0$):
\be  \label{soln}
v(t) = v_0e^{-{\gamma\over m} t} + \frac{1}{m} \int _0^te^{-{\gamma\over m} (t-t')}\xi (t') dt'
\ee
one can obtain $a_2 = \frac{\Gamma}{m^2}$. Thus the Fokker Planck equation becomes:
\be    
{\partial P(v,t)\over \p t} = {\gamma\over m} {\p \over \p v} v P + {\Gamma \over {2m^2}} {\p ^2 P \over \p v^2}
\ee
Finally since we know that $P(v)=e^{-{mv^2\over 2kT}}$ is a time independent solution of the Fokker-Planck equation we get
\be 	\label{FD}
\Gamma = ~{2\gamma k T}
\ee 

This is the fluctuation dissipation  theorem in this context, because it relates $\Gamma$, the strength of the fluctuation, to $\gamma$
the strength of the dissipation. \\

The Langevin equation is much more convenient to work with. To the extent that it assumes that time scales are larger than the microscopic time scale it must fail for very small time scales. As Kubo has shown, stationarity should imply that
\[
{d\over d t_0} \langle v(t_0) v(t_0) \rangle = 0 = \langle \dot v(t_0) v(t_0)\rangle 
\]
Whereas (\ref{soln}) gives 
\[
\langle \dot v(t_0) v(t_0)\rangle = -{\gamma\over m} \langle  v(t_0) v(t_0)\rangle \neq 0
\]
The random force $\xi$ represents the effects of the interaction of other degrees of freedom on our particle and the assumption
that the correlation time is zero is unphysical. A proper microscopic theory that incorporates these effects should not give this contradiction. Kubo has argued that one can replace $\Gamma \delta (t-t')$ by a more general $\Gamma (t-t')$ which
is less singular than a delta function. To see this we modify the Langevin equation to
\be	\label{Langmod}
m\dot v = -\int _{t_0}^t dt' \gamma (t-t') v(t') + \xi (t)
\ee
As long as 
\be \label{delta}
\displaystyle \lim _{t\rightarrow t_0}\int _{t_0}^t dt' \gamma (t-t') v(t')=0 ,
\ee
the aforementioned contradiction is avoided.
Thus
\[
\gamma (\omega) = \int _0^\infty dt~e^{i\omega t} \gamma (t) 
\]
acquires a non trivial frequency dependence. With this it can be shown that
\be	\label{FDmod}
\int _0 ^\infty dt ~e^{i\omega t} \langle \xi(t_0) \xi (t_0+t) \rangle =\Gamma(\omega)= k T \gamma (\omega)
\ee

This is the fluctuation-dissipation theorem that replaces (\ref{FD})\footnote{The factor of 2 has disappeared because in the Laplace transform the integral is from $0$ to $\infty$ and not from $-\infty$ to $\infty$.}.
In fact more generally fluctuation dissipation theorems can be stated in terms of properties of various two point correlation functions.
This is particularly clear in the Schwinger-Keldysh formalism. Son and Teaney have shown how the Schwinger-Keldysh Green functions can be obtained holographically and their
 holographic calculation  gives such a frequency dependent correlation function for the noise which satisfies the FD theorems. However if one expands in powers of frequency one cannot see the softening of the delta function. One needs a more non perturbative result.  \\

In addition to obtaining a Langevin equation for the boundary theory at infinity, Son and Teaney also obtained an effective membrane action and a Langevin equation, at a ``stretched" horizon close to the event horizon. However in AdS$_5$ the equations cannot be solved exactly. Thus the solution had to be worked out as a power series in the frequency.
In this paper we do an almost identical calculation for the case of the BTZ black hole in AdS$_3$ where one can solve the bulk equation of motion exactly. This was first shown in \cite{Mukund} where some exact correlators were computed. We then use the techniques of \cite{Son-Teaney} to obtain the Schwinger-Keldysh Green functions exactly.
We do indeed find the softening of the delta function that avoids the contradiction pointed out by Kubo. It is interesting that internal consistency at the microscopic level is built into the holographic formalism. (Of course the holographic result is in some sense the leading term in a ``strong coupling'' expansion i.e, large N and large $\lambda$ limits of the theory. Departures from large N requires quantum or stringy corrections in the bulk, whereas departure from large $\lambda$ suggests `supergravity' is not a good approximation and needs higher derivative corrections to the gravitational theory. Therefore to ensure consistency at higher orders it may be that one has to embed the boundary theory in a string theory.)


We also find the interesting phenomenon of dissipation at zero temperature.  This is a little puzzling because at zero temperature one expects the system to have Lorentz invariance and boost invariance would say that a quark moving at a constant velocity cannot possibly feel any drag force. One can indeed check in our case, that even though the Green function does have a dissipative component at zero temperature, the frequency dependence is such that force on a constant velocity quark does continue to be zero. Thus there is nothing unphysical about this. Accelerating quarks can certainly experience dissipation by coupling to the massless degrees of freedom in the conformal field theory  - i.e. ``radiation"\cite{Mikhailov,Chernicoff1,Chernicoff2}.  Dissipation at zero temperature has been reported in the literature earlier \cite{Mikhailov,Chernicoff1,Chernicoff2,Gubser2,CT2,Chernicoff,Pitaevskii,Mohanty,Bonart,Cherny,Walmsley,Kamenev}.

We are also able to place the membrane at an arbitrary location without a power series expansion and thus obtain a generalized Langevin equation at an arbitrary location. We believe this may be useful in a holographic RG analysis  of this system.

In the path integral approach to the Langevin equation it is manifest that both $\gamma (\omega)$ and $\Gamma (\omega)$ are related to correlation functions of the noise. $\Gamma$ is related to the symmetric two point function and $\gamma$ to the retarded two point function. The FD theorem is then a statement of a relation between these two correlation functions and what we find is, as expected,
consistent with this theorem. \\

  This paper is organized as follows: Section \ref{sec:LE} is a description of the Langevin equation and its derivation using the Schwinger-Keldysh technique and is a review.  In Section \ref{sec:TS} the retarded Green function is calculated using the usual AdS/CFT prescription. For the BTZ case the Green function can be obtained exactly. This section contains one of the main results of this paper. The Section \ref{sec:SK} is mainly a review where we repeat the Son and Teaney derivation of the Schwinger-Keldysh Green functions using holography. This is also then a verification of the FD theorem. The main point of departure is that the various Green functions that make up the Schwinger-Keldysh Green function are all known exactly in the BTZ case.   Section \ref{sec:EA} starts with a brief review of the holographic RG as discussed in \cite{FLR} and its relevance for our work. It also contains the second main result of this paper in which, by calculating the bulk to bulk propagators exactly, we obtain an effective ``boundary'' action but now with the  boundary at an arbitrary location. From the boundary perspective this is like an effective action at an arbitrary point along the RG flow . In Section \ref{sec:Time scales} different time scales relevant to Brownian motion have been discussed. Section \ref{sec:Conclusions} contains some conclusions.

\section{Langevin Dynamics : A Review}
\label{sec:LE}
Here the Langevin dynamics \cite{Langevin} will be reviewed in brief. Suppose in a viscous medium a very heavy ( compared to the masses of the medium particles ) particle is moving. Its dynamics will be described by the Langevin equation\footnote {This is actually the small-frequency limit of the generalized Langevin equation \eqref{Langmod}. In Section \ref{sec:TS} it will turn out that one obtains the generalized Langevin equation \eqref{Langmod} from holographic calculation rather than its local version \eqref{eq:LE:1}.}
\begin{align}\label{eq:LE:1}
&M_{\text{kin}}\frac{d v}{dt} +\gamma v =\xi(t) \\ \label{eq:LE:2}
\text{with \ \ \ }\langle \xi(t)\xi(t^\prime)\rangle &= \Gamma \delta(t-t^\prime) = 2kT \gamma \delta(t-t')
\end{align}

where $-\gamma v $ is the drag , $\xi$ is the random noise and $M_{\text{kin}}$ is the ``renormalized mass'' in the thermal medium. Evidently equation \eqref{eq:LE:2} is a statement of fluctuation-dissipation theorem. At the ultimate long time limit we can neglect inertial term in \eqref{eq:LE:1} 
\begin{align}\label{eq:LE:3} 
\gamma v =\xi
\end{align}
and can define the diffusion coefficient as 
\begin{align}\label{eq:LE:3.1} 
D=\frac{T}{\gamma} 
\end{align}

We will see later the dynamics on the stretched horizon \eqref{eq:damped} is identical to this overdamped motion \eqref{eq:LE:3}. 

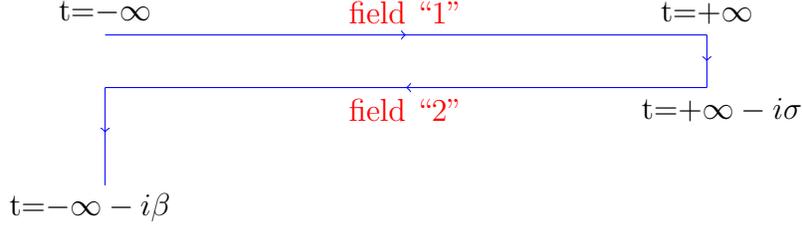
\begin{figure}
\begin{center}
\begin{tikzpicture} 
\draw [->,blue] (-4,0) -- (0,0);
\draw [blue] (0,0) -- (4,0);
\draw [->,blue] (4,-.7) -- (0,-.7);
\draw [blue] (0,-.7)--(-4,-.7);

\draw [->,blue] (4,0) -- (4,-.35);
\draw [blue]  (4,-.35)--(4,-.7);
\draw [->,blue] (-4,-.7) -- (-4,-1.3);
\draw [blue](-4,-1.3)--(-4,-2);
\draw (-4,0.3) node{t=$-\infty$};
\draw (4,0.3) node{t=$+\infty$};
\draw (4.2,-1) node{t=$+\infty-i\sigma$};
\draw (-4.2,-2.3) node{t=$-\infty-i\beta$};
\draw [red](0,.3) node{field ``1''};
\draw [red](0,-1) node{field ``2''};
\end{tikzpicture}
\caption{\emph{Schwinger-Keldysh contour for systems in thermal equilibrium with temperature $\beta^{-1}$ }} \label{fig:S-K}
\end{center} 
\end{figure}

The aim of this section is to review how to derive Langevin equation from path integral formalism. There are many good references \cite{GSI,GL} for detail description of this derivation, we will go through this quickly just to fix the notation we will use through out this article and we follow mostly the steps sketched in \cite{Son-Teaney}. We can define the partition function for a heavy particle in a heat bath using a Schwinger-Keldysh contour (fig.\ref{fig:S-K})
\hspace{4cm}

\begin{align}\label{eq:LE:4}
Z= \left \langle \int [\mathcal{D}x_1] [\mathcal{D}x_2] ~ e^{i\int dt_1 M_Q^0 \dot{x}_1^2} ~e^{-i\int dt_2 M_Q^0 \dot{x}_2^2} ~e^{i\int dt_1 \phi_1(t_1)x_1(t_1)} ~e^{-i\int dt_2 \phi_2(t_2)x_2(t_2)}\right \rangle_{\text{bath}} 
\end{align}
$\phi_1$, $\phi_2$ are the heat bath degrees of freedom which act like sources. $x_1 , x_2$ are the fields ``living'' on the two different sections 1 and 2 of the time contour. We will see later in the gravity side these are the two types of field those ``live'' on the two boundaries, 1 and 2 of the full Kruskal diagram (fig.\ref{fig:Kruskal}). Path integral along the vertical portion of the contour gives us average over the thermal density matrix $e^{-\beta H}$. And $\sigma$ is a free parameter that can take any value. For this discussion we can safely choose $\sigma = 0$ and later we will see that this choice is necessary for ``$ra$ formalism'' which is used extensively in real time thermal field theory literature \cite{CSHY,HM,GL,Caron-Huot}. \\

For very heavy particle we can consider the forces on the particle is very small compared to inertial term so we can expand it in second order , take the average over bath and make it an exponentiate again to get 
\begin{align}\label{eq:LE:5}
Z= \int [\mathcal{D}x_1] [\mathcal{D}x_2] ~ e^{i\int dt_1 M_Q^0 \dot{x}_1^2} ~ e^{-i\int dt_2 M_Q^0 \dot{x}_2^2}~ e^{-\frac{1}{2}\int dt dt^{\prime}x_s(t)[\langle \phi(t)\phi(t^{\prime}) \rangle]_{ss^{\prime}}x_{s^{\prime}}(t^{\prime})} 
\end{align} 
Here the Green function takes a $2\times2$ matrix form as there are two type of fields and it is contour ordered .
\begin{align}\label{eq:LE:6}
[\langle \phi(t)\phi(t^{\prime}) \rangle]_{ss^{\prime}}\equiv i \begin{pmatrix} G_{11}(t,t^{\prime}) & -G_{12}(t,t^{\prime})\\ -G_{21}(t,t^{\prime}) & G_{22}(t,t^{\prime})\end{pmatrix}
\end{align}
Notice that $G_{11}(t,t^{\prime})$ is the usual time ordered Feynman Green function where as $G_{22}(t,t^{\prime})$ is anti-time ordered Green function. \\

In operator language, if we define 
\be \label{phi}
\phi(t)=e^{iHt}\phi(0)e^{-iHt}
\ee
The different Green functions are defined as, 
\begin{align}
i G_{11}(t,t^\prime)&=\langle \mathcal {T} \phi(t) \phi(t^\prime) \rangle \\
i G_{22}(t,t^\prime)&=\langle \tilde{\mathcal {T}} \phi(t) \phi(t^\prime) \rangle \\
i G_{12}(t,t^\prime)&=\langle  \phi(t^\prime)\phi(t) \rangle \\
i G_{21}(t,t^\prime)&=\langle \phi(t) \phi(t^\prime) \rangle 
\end{align}

The KMS relation $Tr[e^{-\beta H}\phi(t)\phi(0)]=Tr[e^{-\beta H}\phi(0)\phi(t+i\beta)]$ is easy to prove using cyclicity of the trace and the definition (\ref{phi}). This implies for the Fourier transform
\begin{align} \label{KMS}
e^{\beta \omega} \int_{-\infty}^{\infty} dt~ e^{i\omega t} \langle Tr[e^{-\beta H}\phi(0)\phi(t)]\rangle = \int_{-\infty}^{\infty} dt ~e^{i\omega t}Tr[e^{-\beta H} \phi(t)\phi(0)]
\end{align}

In addition to this
if we add $iG_{11}+iG_{22}$ we will get 
\begin{align}
iG_{11}(t,t^\prime)+iG_{22}(t,t^\prime)=&\langle \mathcal {T} \phi(t) \phi(t^\prime) \rangle + \langle \tilde{\mathcal {T}} \phi(t) \phi(t^\prime) \rangle \nonumber \\
=&\langle \phi(t) \phi(t^\prime) \rangle \{\theta(t-t^\prime)+\theta(t^\prime-t) \} \nonumber \\
 \hspace{3cm}&+\langle \phi(t^\prime) \phi(t) \rangle \{\theta(t^\prime-t)+ \theta(t-t^\prime)\} \nonumber \\
=&i G_{12}(t,t^\prime)+i G_{21}(t,t^\prime)
\end{align}
Therefore we can write 
\begin{align}   \label{G}
G_{11}+G_{22}=G_{12}+G_{21}
\end{align}

Note that \eqref{G} is true only for $\sigma=0$. We will see our Green functions will obey this relation.

Using (\ref{KMS}) and (\ref{G})
all these components of the matrix can be expressed in terms of any one Green function, say retarded Green function
\begin{align}\label{eq:LE:7}
 i~ G_R(t)= \theta(t)\left \langle [\phi(t),\phi(0)]\right \rangle_{\text{bath}}
\end{align}
Thus for instance we can use
\be
\text{Im}~G_R(\omega) = -\frac{i}{2}\int _{-\infty}^\infty dt~ e^{i\omega t}\langle [\phi(t),\phi(0)]\rangle =-\frac{ i(e^{\beta\omega}-1)}{2}\int _{-\infty}^\infty dt~ e^{i\omega t}\langle \phi(0)\phi(t)\rangle
\ee

to write
\be   \label{FD}
\text{Im}~ G_R(\omega) = -i~ \text{tanh} \left({\beta \omega\over 2}\right) G_{\text {sym}}(\omega)
\ee
where\footnote{\text{According to our definition} $G_{\text {sym}}(\omega)$ \text{is purely imaginary.} } $G_{\text {sym}}(\omega)= {1\over 2}\int _{-\infty}^\infty dt~\langle \{\phi(t),\phi(0)\}\rangle e^{i\omega t}$.  \\
\par
Now we will introduce the previously advertised \emph {ra} formalism. We have already taken $\sigma=0$ . As we are working with very heavy quark the motion will be nearly classical. So, $x_1$ $\sim x_2$ . Therefore we can use some sort of ``centre of mass'' coordinates for the particle and for the forces too.
\begin{align}\label{eq:LE:8}
x_r=\frac{x_1+x_2}{2} \hspace{3cm} x_a=x_1-x_2
\end{align}\begin{align}\label{eq:LE:8.1}
\phi_r=\frac{\phi_1+\phi_2}{2} \hspace{3cm} \phi_a=\phi_1-\phi_2
\end{align}
$r$ and $a$ here refer to retarded and advanced respectively and we should remember $x_a$ is a very small quantity for quasi-classical description. Now substituting \eqref{eq:LE:8} and \eqref{eq:LE:8.1} into the partition function \eqref{eq:LE:5} 
\begin{align}\label{eq:LE:9}
Z= \int [\mathcal{D}x_r] ~[\mathcal{D}x_a] ~ e^{-i\int dt M_Q^0 x_a \ddot{x}_r} ~ e^{-i\int dt dt^\prime [x_a(t)iG_R(t,t^\prime) x_r(t^\prime)-\frac{1}{2}x_a(t)G_{\text{sym}}(t,t^\prime) x_a(t^\prime)]}
\end{align}

where the propagators 
\begin{align}\label{eq:LE:10}
&G_{\text{sym}}(t,t^\prime)=\langle \phi_r(t)\phi_r(t^\prime)\rangle = \frac{1}{2}\left \langle \{\phi(t),\phi(t^\prime)\}\right \rangle \\ \label{eq:LE:11}
&iG_R(t,t^\prime)=\langle \phi_r(t)\phi_a(t^\prime)\rangle = \theta(t-t^\prime)\left \langle [\phi(t),\phi(t^\prime)]\right \rangle
\end{align}

As we have argued earlier the different Green functions are not independent. In particular the retarded and the symmetric Green function are related .
\begin{align}\label{eq:LE:12}
iG_{\text{sym}}(\omega) = -(1+2n_B) ~\text{Im}~G_R(\omega) 
\end{align} 

Here $n_B(\omega) = \frac{e^{-\beta \omega}}{1-e^{-\beta \omega}}$ is the Bosonic occupation number . This is just a rewriting of \eqref {FD}.\\

This is a canonical statement of fluctuation-dissipation theorem. Later in this section we will identify these two Green functions as $\gamma(\omega)$ and $\Gamma(\omega)$ of the equations \eqref{Langmod} and \eqref{FDmod}. Now we can write down the path integral in Fourier space 
\begin{align}\label{eq:LE:13}
Z= \int [\mathcal{D}x_r]~ [\mathcal{D}x_a]  ~\text{exp}\left(-i\int \frac{d\omega}{2\pi} x_a(-\omega)[ - M_Q^0 \omega^2 + G_R(\omega)]x_r(\omega)\right) e^{-\frac{1}{2}\int \frac{d\omega}{2\pi} x_a(-\omega)[iG_{\text{sym}}(\omega)]x_a(\omega)}
\end{align}
Now we introduce a new random variable which we call $\xi$ in anticipation that it will turn out to be the random noise, by defining 
\begin{align}\label{eq:LE:14}
e^{-\frac{1}{2}\int \frac{d\omega}{2\pi} x_a(-\omega)[iG_{\text{sym}}(\omega)]x_a(\omega)}= \int [\mathcal{D}\xi]~ e^{i \int x_a(-\omega) \xi(\omega)} e^{-\frac{1}{2}\int\frac{\xi(\omega)\xi(-\omega)}{iG_{\text{sym}}(\omega)}\frac{d\omega}{2\pi}}
\end{align}
The partition function becomes
\begin{align}\label{eq:LE:15}
Z= \int [\mathcal{D}x_r]~ [\mathcal{D}x_a]~ [\mathcal{D}\xi] ~e^{-\frac{1}{2}\int \frac{d\omega}{2\pi}\frac{\xi(\omega)(-\omega)}{iG_{\text{sym}}(\omega)}}~ \text{exp}\left(-i\int \frac{d\omega}{2\pi} x_a(-\omega)[ - M_Q^0 \omega^2x_r(\omega) + G_R(\omega)x_r(\omega)-\xi(\omega)]\right)
\end{align}
Integrate out $x_a(-\omega)$ to get a delta function in $\omega$-space
\begin{align}\label{eq:LE:16}
Z= \int [\mathcal{D}x_r] ~[\mathcal{D}\xi] ~e^{-\frac{1}{2}\int \frac{d\omega}{2\pi}\frac{\xi(\omega)(-\omega)}{iG_{\text{sym}}(\omega)}} ~\delta_\omega \left[- M_Q^0 \omega^2x_r(\omega) + G_R(\omega)x_r(\omega)-\xi(\omega)\right]
\end{align}
This partition function is an average over the classical trajectories for the heavy particle under the noise $\xi$ . 
\begin{align}\label{eq:LE:17}
\left[- M_Q^0 \omega^2 + G_R(\omega)\right] x(\omega)= ~\xi(\omega) \hspace{2cm} \langle\xi(-\omega)\xi(\omega) \rangle = ~i G_{\text{sym}}(\omega)
\end{align}
Going back to time space we obtain the generalized Langevin equation 
\begin{align}\label{eq:LE:18}
M_Q^0\frac{d^2x(t)}{dt^2}+ \int_{-\infty}^t dt^\prime ~G_R(t,t^\prime)x(t')=~\xi(t) \hspace{1.7cm} \langle\xi(t)\xi(t^\prime) \rangle =~ i G_{\text{sym}}(t,t^\prime)
\end{align}
$G_R(t,t')$ is thus the same as $\gamma(t-t')$ of  \ Section \ref{sec:intro} for the choice $t_0=-\infty$ and $iG_\text{sym}(t,t')$ is the same as $\Gamma(t-t')$.

Now if the Green function is expanded for small frequencies the coefficient of $\omega^2 \left ( i.e, \frac{d^2x(t)}{dt^2} \right)$  adds to the mass of the particle and the coefficient of $\omega \left( i.e,\frac{dx(t)}{dt}\right)$ will contributes as the drag term 
\begin{align}\label{eq:LE:19}
G_R(\omega)=-\Delta M \omega^2- i \gamma \omega + \ldots
\end{align}
After taking into account the thermal mass correction we define the effective mass $$ M_{\text{kin}}(T)= M_Q^0 + \Delta M $$ Then the Langevin equation reads 
\begin{align}\label{eq:LE:20}
M_{\text{kin}}\frac{d^2x}{dt^2}+\gamma \frac{dx}{dt}=~\xi \\\label{eq:LE:20}
\text{with \ \ \ }\langle \xi(t)\xi(t^\prime)\rangle =~\Gamma(t-t^\prime)
\end{align}
These equations are identical to \eqref{eq:LE:1} and \eqref{eq:LE:2}.

\section{Obtaining the Generalized Langevin Equation from Holography}
\label{sec:TS}

The Einstein-Hilbert action for AdS$_3$  which has negative cosmological constant, $-~\Lambda = \frac{1}{L^2}$  is given by
\begin{align}
\text{I}_{\text{EH}}= ~ \frac{1}{2\pi} \int ~\mathrm d{t}~ \mathrm d {r}~ \mathrm d{x} ~\sqrt{-g}~\left[R + \frac{2}{L^2}\right] +~ \text{I}_{\text{B'dy}}
\end{align}
In this units $16\pi G$ is same as $2\pi$. And therefore, $8G=~1$. Since $G$ goes as \emph{length} in 2+1 dimensions this defines a choice of length units. 

We write down the action for a string stretching from the horizon ($r=r_h$) towards the AdS boundary and ending on the probe brane at $r=r_m$ , in background metric of $\text{AdS}_3$ with a BTZ  black hole embedding. 
The BTZ metric is\begin{align}\label{metric2}
ds^2=-\left(\frac{\bar{r}^2}{L^2}-8GM\right)\mathrm d t^2+{\left(\frac{\bar{r}^2}{L^2}-8GM\right)}^{-1}\mathrm d \bar{r}^2+\frac{\bar{r}^2}{L^2}\mathrm d x^2
\end{align}

Let's write this background metric as
\begin{align} \label{metric1}
ds^2=\frac{\bar{r}^2}{L^2}\left[-f(b\bar{r})\mathrm d t^2+ \mathrm d x^2 \right]+\frac{L^2 \mathrm d \bar{r}^2}{f(b\bar{r})\bar{r}^2}
\end{align}
where $\bar{r}$ is the canonical choice of coordinate with dimension of length, $b$ is the inverse horizon radius, $L$ is the AdS radius. 
In our unit, $r_h=~b^{-1}=~\sqrt{8GM}~L$ . 
\\ So, $f(b\bar{r})=1-\frac{8GML^2}{r^2}$. Thus $f(r)=1-{1\over r^2}$ and  $\pi T=~\frac{\sqrt{8GM}}{2L}=~\frac{\sqrt{2GM}}{L}$ defines the Hawking temperature.
$ b=\frac{1}{2 \pi T L^2}$ is an alternate expression for $b$, which can be taken to be the black hole mass parameter.

 Now we will write the same metric \eqref{metric1} with a dimensionless coordinate, $ r\equiv b\bar{r}$ 
\begin{align}\label{metric3}
ds^2=(2 \pi T)^2 L^2 \left[-r^2 f(r) \mathrm d t^2+r^2 \mathrm d x^2 \right]+\frac{L^2 \mathrm d r^2}{r^2f(r)}
\end{align}

We want to study the small fluctuation of the string in this non trivial background. The Nambu-Goto action is  
\begin{align}
S=- \frac{1}{2 \pi l_s^2 }\int \mathrm d{\tau} \mathrm d {\sigma} ~\sqrt{-\det h_{ab}}
\end{align}
Target space coordinates are, $$X^{\mu}\equiv (t,r,x)$$ 
And world sheet coordinates are, $$\sigma_0 = \tau \text{\hspace{4mm} and \hspace{4mm}} \sigma_1 = \sigma $$
Now we will choose (static gauge), $$ t = \tau \text{\hspace{4mm} and \hspace{4mm}} r = \sigma $$
So, $$x\equiv x(\tau,\sigma)=x(t,r)$$ \\ 

Induced metric , $$ h_{ab}= ~G_{\mu \nu} \frac{\mathrm d {X^{\mu}}}{\mathrm d {\sigma_a}}  \frac{\mathrm d {X^{\nu}}}{\mathrm d {\sigma_b}}\hspace{1cm} a,b=0,1 $$  \\

$G_{\mu \nu}$ is the target space metric which is $\text{AdS}_3$-BH for present case. \\

\[ h \equiv \det(h_{ab}) = \left| \begin{array}{cc}
G_{\mu \nu} \frac{\mathrm d {X^{\mu}}}{\mathrm d {\tau}}  \frac{\mathrm d {X^{\nu}}}{\mathrm d {\tau}} & ~~G_{\mu \nu} \frac{\mathrm d {X^{\mu}}}{\mathrm d {\tau}}  \frac{\mathrm d {X^{\nu}}}{\mathrm d {\sigma}}\\ \\
G_{\mu \nu} \frac{\mathrm d {X^{\mu}}}{\mathrm d {\sigma}}  \frac{\mathrm d {X^{\nu}}}{\mathrm d {\tau}}& ~~G_{\mu \nu} \frac{\mathrm d {X^{\mu}}}{\mathrm d {\sigma}}  \frac{\mathrm d {X^{\nu}}}{\mathrm d {\sigma}}  \end{array} \right|\]
\vspace{.5cm}
\[\hspace{2.45 cm}= \left| \begin{array}{cc}
G_{tt}+G_{xx}\dot{x}^2  & G_{xx}\dot{x}x^{\prime}\\ \\
G_{xx}x^{\prime}\dot{x}& G_{rr}+G_{xx}{x^{\prime}}^2 \end{array} \right|\] 

\begin{align*}
h= -(2\pi T)^2 L^4 \left[1+ (2\pi T)^2 r^4 f(r){x^{\prime}}^2 -\frac{\dot{x}^2}{f(r)}\right ]
\end{align*}  
For small fluctuations $x^{\prime}$ and $\dot{x}$ are very small. So we can write  
\begin{align*}
\sqrt{-h}&= (2\pi T) L^2 \sqrt{1+ (2\pi T)^2 r^4 f(r){x^{\prime}}^2 -\frac{\dot{x}^2}{f(r)}} \\
&\approx (2\pi T) L^2 \left [1+ \frac{1}{2}(2\pi T)^2 r^4 f(r){x^{\prime}}^2 -\frac{1}{2}\frac{\dot{x}^2}{f(r)}\right ]
\end{align*}

Action for the small fluctuation of string world sheet 
\begin{align*}
S= - \frac{(2\pi T) L^2 }{2 \pi l_s^2 }\int \mathrm d t \mathrm d r \left [1+ \frac{1}{2}(2\pi T)^2 r^4 f(r){x^{\prime}}^2 -\frac{1}{2}\frac{\dot{x}^2}{f(r)}\right ]
\end{align*}

Now define , mass per unit $r$
\begin{align}\label{Mass}
m \equiv \frac{(2\pi T) L^2 }{2 \pi l_s^2 }=\sqrt{\lambda}T ~~;~\hspace{.6cm} \text{ defining,~~} {\left (\frac{L}{l_s}\right )}^4 \equiv \lambda
\end{align}

And the local tension 
\begin{align}\label{Tension}
T_0(r)\equiv \frac{(2\pi T)^3 L^2 }{2 \pi l_s^2 } f r^4 = 4\sqrt{\lambda}\pi^2 T^3 f r^4 = 4\sqrt{\lambda}\pi^2 T^3 r^2(r^2-1)
\end{align}

Then the action reduces to 
\begin{align} \label{Action}
S= -\int \mathrm d t \mathrm d r \left [m + \frac{1}{2}T_0 {(\partial_r x)}^2 -\frac{m}{2f}{(\partial_t x)}^2\right ]
\end{align}

The equation of motion (EOM) can be obtained by varying the action ($ \delta S = 0$ ), 
\begin{align}
0= -\frac{m}{f}\partial_t^2 x + \partial_r (T_0(r)\partial_r x)
\end{align}

Then the standard way is to write down the EOM in Fourier space
\begin{align*}
x(r,t)&= \int \frac{\mathrm d \omega}{2 \pi } e^{i \omega t}f_\omega(r)x_0(\omega) \\ 
x(r=r_m,t)&= \int \frac{\mathrm d \omega}{2 \pi } e^{i \omega t}x_0(\omega) \hspace{2cm} \text{with,}\hspace{0.4 cm}f_\omega(r_m)=1
\end{align*}

Therefore the EOM in terms of the modes reduces to 
\begin{align} \label{EOM}
\frac{{\mathfrak {w}}^2}{f}f_\omega(r) + \partial_r[fr^4 \partial_r f_\mathfrak\omega(r)] = 0 
\end{align}
where we have defined $\mathfrak{w}\equiv \omega/(2\pi T)$ . \\ 

For our case $f(r)=1-\frac{1}{r^2}$, so the EOM reduces to
\begin{align}\label{eq:TS:EOM}
\partial_r^2 f_\omega + \frac{2(2r^2-1)}{r(r^2-1)}\partial_rf_\omega + \frac{\mathfrak{w}^2}{(r^2-1)^2} f_\omega =0 
\end{align}
This is an ordinary second order linear differential equation in $r$. This can be recast into associated Legendre differential equation which one can solve exactly\footnote{The exact solution to this EOM for a stochastic string in BTZ background was obtained earlier by J. de Boer et al. in \cite{Mukund} to calculate some exact correlators.}. The general solution to the EQM will be 
\begin{align}
f_\omega(r)= ~C_1 ~\frac{P^{i\mathfrak{w}}_1}{r} + C_2 ~\frac{Q^{i\mathfrak{w}}_1}{r}
\end{align}

 where $P^\mu_\lambda$ and $Q^\mu_\lambda$ are associated Legendre functions and $C_1$,$C_2$ are two constants which will be determined by two boundary conditions at the horizon ($r=1$) and at the boundary  ($r\to \infty$) of the AdS space. We will impose the following boundary conditions on the modes, $f_\omega$ to obtain the retarded Green function as prescribed by Son and Starinets \cite{SS}.  Actually this ``prescription'' has been derived quite rigorously later by van Rees in \cite{Rees} based on the dictionary of the Lorentzian AdS/CFT as formulated in \cite{Rees-Skenderis}.  Furthermore the application of this formalism to holographic Brownian motion is described in Appendix D of  \cite{AJM} .  \\
 
1. At the \textit{horizon} to impose the ingoing wave boundary condition one has to pick the solution  $\frac{P^{i\mathfrak{w}}_1}{r}$ ( see appendix \ref{sec:A}).
So, $$ f^R_\omega(r) \sim \frac{P^{i\mathfrak{w}}_1}{r} $$
  
2. The other condition that it should satisfy at the ``\textit{boundary}'' ( $r_m$ , say, where $r_m >> 1$) of the AdS space is, $r \to r_m $ , $f^R_\omega(r) \to 1 $ . 
\begin{align}
f^R_\omega(r)&= \frac{(1+r)^{i\mathfrak{w}/2}}{(1+r_m)^{i\mathfrak{w}/2}} ~\frac{(1-r)^{-i\mathfrak{w}/2}}{(1-r_m)^{-i\mathfrak{w}/2}} ~ \frac{r_m}{r}~ \frac{{}_2F_1(-1,2;1-i\mathfrak{w};\frac{1-r}{2})}{{}_2F_1(-1,2;1-i\mathfrak{w};\frac{1-r_m}{2})} \nonumber \\\label {eq:TL:mode}
&=  \frac{(1+r)^{i\mathfrak{w}/2}}{(1+r_m)^{i\mathfrak{w}/2}} ~\frac{(1-r)^{-i\mathfrak{w}/2}}{(1-r_m)^{-i\mathfrak{w}/2}} ~ \frac{r_m}{r}~ \frac{\mathfrak{w} +i r}{\mathfrak{w} +i r_m }
\end{align} 

Now the retarded correlator $ G_R(\omega)$ is defined as 
\begin{align} \label {RGF}
G^0_R \equiv \lim_{r \to r_m } T_0(r)f^R_{-\omega}(r) \partial_r f^R_{\omega}(r)= ~-~M^0_Q \omega^2 + G_R(\omega)
\end{align} 

$ M^0_Q $ is zero temperature mass of the external particle and the term containing it comes from the ``divergent part'' of the boundary limit ( i.e, $r_m \to \infty$). Our goal here is to extract $ G_R(\omega)$ and then some interesting physical quantities like viscous drag and mass shift form it. \\

\par $r_m$ is a regulator here. So to calculate the retarded correlator we should take the limit $ r \to r_m $. Taking this limit, from \eqref{eq:TL:mode},

\begin{align}
\partial_r f^R_{\omega}(r)\bigg|_{r\to r_m}= ~-~ \frac{\mathfrak{w} (r_m \mathfrak{w} +i)}{r_m ({r_m}^2 -1) (r_m -i \mathfrak{w} )}
\end{align}

and using the fact that $f^R_{-\omega}(r_m)=1 $ ,  we obtain

\begin{align}
G^0_R &= T_0(r)f^R_{-\omega}(r) \partial_r f^R_{\omega}(r) \bigg{|}_{r\to r_m} \nonumber \\ \label {eq:TL:GF1a}
&=~-~ 4 \sqrt{\lambda}\pi^2 T^3 ~ \frac{ r_m \mathfrak{w} (r_m \mathfrak{w} +i)(r_m +i \mathfrak{w} )}{(r_m^2 +\mathfrak{w}^2 )} \\ \label {eq:TL:GF1b}
&=~-~ 4 \sqrt{\lambda}\pi^2 T^3~ \frac{ r_m \mathfrak{w} (r_m \mathfrak{w} +i)}{(r_m -i \mathfrak{w} )} 
\end{align}

Equation \eqref{eq:TL:GF1b} is an exact expression for the retarded force-force correlator for the boundary field theory. Note also that it has a singularity only in the lower half $\omega$-plane as required for a retarded Green function.
But it is written in terms of two dimensionless parameters $r$ and $\mathfrak{w}$. To make the scaling behavior of the boundary theory correlator more natural we use the corresponding dimensionful parameters, namely $\omega= 2\pi T\mathfrak{w}$ , $\bar{r}_m= 2\pi T L^2 r_m$. \\ 
Now we can introduce a mass scale $\mu\equiv \frac{\bar{r}_m}{l_s^2}$. Actually, as discussed in Section \ref {sec:HRG},  $\mu$ can be treated as a renormalization group (RG) scale for the dual field theory. Therefore we do not have to push $\mu$ all the way to infinity\footnote{There is also another physical reason why it shouldn't be pushed all the way to the boundary. In such case the mass of the heavy quark is infinite and there would be no Brownian motion.}. We would rather take the point of view that the parameters of the field theory run with $\mu$ such a way that the physical quantities remain unchanged. So the correlator reduces to  
\begin{align}\label {eq:TL:GF1.1}
G^0_R(\omega) &= ~-\mu \omega ~{(i \sqrt \lambda~ 4 \pi ^2 T^2 + \mu \omega )\over 2\pi  ( \mu - i \sqrt \lambda \omega )} 
\end{align} 

Now absorbing the divergent piece ( the leading term in the large $\mu$ expansion  which goes as $\mu \omega^2$ ) in the definition of the zero temperature mass of the Brownian particle and subtracting it from $G^0_R$ we can define the retarded boundary Green function, $G_R$ 
\begin{align}\label {eq:TL:GF2}
G^0_R \equiv ~-~M^0_Q \omega^2 + G_R(\omega)  
\end{align} 
\begin{align}  \label{eq:Mass}
\text{where,}~~~ M^0_Q &=\frac{\mu}{2 \pi} \\  \nonumber
&= \sqrt{\lambda} T ~r_m \\ \nonumber &={\sqrt{\lambda}\over \pi L^2}\bar r_m 
\end{align} 
 As mentioned above, $M_Q^0$ is nothing but the mass of the string stretching from $\bar {r}_m$ to $0$ in the zero temperature limit.
And, 
\begin{align}\label{eq:TL:GF3}
G_R(\omega)&= ~-~ {i \sqrt \lambda \mu \omega  (4 \pi ^2 T^2 + \omega^2)\over 2 \pi (\mu - i \sqrt \lambda \omega )}\\\label{eq:TL:GF3.1}
&= ~~ {\mu \omega \over 2\pi}~{(\omega^2 + 4 \pi ^2 T^2)\over (\omega + i {\mu\over \sqrt \lambda})}
\end{align} 

$G_R(\omega)$ is clearly finite in the $\mu \rightarrow \infty$ limit.

Now expanding $G_R$ \eqref{eq:TL:GF3} in small frequencies , $\omega$ 
\begin{align}\label{eq:TL:GF4}
G_R(\omega) \approx \frac{2\lambda \pi T^2}{\mu} \omega^2 -i\left(2\sqrt{\lambda}\pi T^2 \omega +\left(\frac{\sqrt{\lambda}}{2 \pi}-  {2(\sqrt \lambda)^3 \pi T^2 \over  \mu^2}\right) \omega^3 \right)
\end{align}

Again we know when $G_R(\omega)$ is expanded in small $\omega$ it takes the form
\begin{align}\label{eq:TL:GF5}
G_R(\omega)= -i\gamma ~\omega - \Delta M \omega^2 -i \rho ~\omega^3+ \ldots 
\end{align}

where $\gamma$ and $\Delta M$ are the viscous drag and the thermal mass shift for the Brownian particle. Whereas $\rho$ is some higher order ``drag coefficient'' as it is known that the imaginary part of the retarded Green function (Im G$_R(\omega)$) is responsible for dissipation. \\

Now comparing \eqref{eq:TL:GF4} and \eqref{eq:TL:GF5} we can identify 
\begin{align} \label{eq:TL:GF6}
\gamma = 2 \sqrt{\lambda}\pi T^2
\end{align}

\begin{align}\label{eq:TL:GF7}
\Delta M &=-~\frac{2 \lambda \pi T^2}{\mu} \\ 
&= -~\sqrt{\lambda}T~\frac{1}{r_m} \\ \nonumber\\ \label{eq:TL:GF8}
\rho&= \frac{\sqrt{\lambda}}{2\pi} -{2(\sqrt \lambda)^3\pi T^2 \over \mu^2}
\end{align}

Note that the particle's rest mass at zero temperature, $ M^0_Q $ \eqref{eq:Mass} and its viscous drag, $ \gamma $ \eqref{eq:TL:GF6} are identical to that of a quark in an $\mathcal{N}=4$ SYM plasma at finite temperature in 3+1 dimensions \cite{HKKKY,CT,Gubser,Son-Teaney}. But the thermal mass shift, $\Delta M$ is vanishingly small for large  value of $\mu$. We have intentionally kept the $\mathcal{O}(r_m^{-1})$ term to look at its leading behavior. $\Delta M $ has the correct dimension  since $r_m$ is a dimensionless quantity ( $r_m=b \bar{r}_m$ and $b\sim length^{-1}$ and $\bar{r}_m \sim length $). \\

One can also compare this mass shift with that obtained by considering the change in mass of a {\em static} string coming from the change in its length due to the presence of a horizon.

\begin{align*}
\Delta M &= - \displaystyle \int_0^{\bar{r}_h} (\text{Tension}) . \sqrt{- g}~ d\bar{r} \\
&= - \frac{1}{2 \pi l_s^2} \int_0^{\bar{r}_h} \sqrt{- g_{tt} g_{rr}}~ d\bar{r} \\
&= - \frac{\bar{r}_h}{2 \pi l_s^2} \\
&= - \frac{T L^2}{l_s^2} \\
&= - \sqrt{\lambda} T
\end{align*} 

One sees that it is not quite the same as \eqref{eq:TL:GF7}. In lower dimension systems the effect of fluctuations could be much larger and could explain the discrepancy.

Notice that if we  take the limit $\mu \to \infty$
 (ultra-violet  limit) and $T \to 0$ we obtain 
\begin{align} \label{CFT}
G_R(\omega)&= -~i \frac{\sqrt{\lambda}}{2 \pi}\omega^3 
\end{align}
which doesn't contain any dimensionful parameter other than $\omega$ and thus properly describes a conformal field theory at UV fixed point. \\

We can now put $T=0$ in \eqref{eq:TL:GF3} to get

\begin{align} \label{eq:zerotemp}
G_R(\omega)\bigg{|}_{ T=0}=~{\mu \omega^3 \over 2\pi (\omega + i {\mu\over \sqrt \lambda})} = ~{\mu \omega^3 (\omega - i {\mu\over \sqrt \lambda})\over 2\pi (\omega ^2+  ({\mu\over \sqrt \lambda})^2)}
\end{align}

The presence of an imaginary part in $G_R(\omega)$ signifies dissipation.
Thus  an interesting result we get from the expression \eqref{eq:zerotemp}  is a temperature independent dissipation. \\
\begin{itemize}
\item {For low frequency regime ($\omega<< \mu$) at zero temperature we recover \eqref{CFT} which shows the diffusive behavior,
\begin{align}\label{small freq}
G_R(\omega)\bigg{|}_{ T=0} \approx -~i \frac{\sqrt{\lambda}}{2 \pi}\omega^3 
\end{align}} 

\item {If we consider the  frequency range $\omega >>\mu$  \footnote{If we are to think of $\mu$ as an effective cutoff of the theory, then we should keep
$\omega < \mu$. So this is only a formal limit.}
\begin{align}\label{large freq}
G_R(\omega)\bigg{|}_{ T=0} \approx ~~{\mu \omega^2\over 2\pi} - i {\mu^2 \omega \over 2 \sqrt \lambda \pi} + \ldots 
\end{align}
We see a drag like term proportional to $\omega$. This term strongly suggests that there must be ``drag'' for the heavy particle even at zero temperature for this 1+1 d CFT.} \\
\end{itemize}

 At first sight this is puzzling because Lorentz invariance of a theory would say that a quark moving with a constant velocity for {\em all} time, should not slow down - this would violate boost invariance\footnote{A Similar phenomenon has been observed for theories with hyperscaling violation \cite{Tong,Pedraza1}. Clearly these backgrounds break Poincare invariance. For these non-relativistic situations, energy and momentum are conserved but drain into the soft infra-red modes of the theory \cite{Tong,Kiritsis}. Moreover, this mechanism of energy loss is present even at {\em constant velocity} of the particle! Evidently this phenomenon is quite different from the one we are addressing in the present article. }. In fact the drag force on a particle moving with a constant velocity turns out to be zero as we see below. The drag force $F(t)$ is given by (in frequency space)
\[
F(\omega)= G_R(\omega)x(\omega)
\]
For a particle moving at constant velocity $x(t)=vt$. This translates to
\[
x(\omega) = -i v \delta ' (\omega)
\]
Since $G_R(\omega =0)=G_R'(\omega=0)=0$, the force is zero. In more detail, since we have a distribution $\delta '(\omega)$,  we should consider a smooth function $f(\omega)$ and evaluate the integral:
\[
\int \mathrm d\omega ~G_R(\omega)x(\omega)f(\omega) =\int \mathrm d\omega ~G_R(\omega)(-i v \delta ' (\omega)) f(\omega)=0
\] on integrating by parts.

(One can trace this to  \eqref{eq:zerotemp} which says that $G_R(\omega)$ starts off as $\omega^3.)$ \\

The phenomena of zero temperature dissipation have been noticed in holography \cite{Mikhailov,Chernicoff1,Chernicoff2,Gubser2,CT2,Chernicoff} and many other contexts \cite{Pitaevskii,Mohanty,Bonart,Cherny,Walmsley,Kamenev}. The physical mechanism that gives rise to energy loss at zero temperature in the relativistic theories was first explained in \cite{Mikhailov}, and then elaborated on in \cite{Chernicoff1,Chernicoff2}. For accelerated quarks in the vacuum of a CFT, energy and momentum are radiated away by emission of gluonic quanta (SYM fields), in analogy with the theory of radiation in classical electrodynamics. This in turns leads to a Lienard-like formula for the rate of energy loss and a generalized Lorentz-Dirac equation that captures the effects of radiation damping. The previous interpretation agrees with the fact that a quark moving with constant velocity does not feel drag force and thus, does not radiate. Moreover, for the Langevin dynamics around accelerated trajectories at zero temperature \cite{Xiao,Pedraza3}, it has also been seen that the stochastic motion of the heavy probe is not due to collisions with the fluid constituents but rather arises due to the random emission of gluonic radiation\footnote{ This interpretation is further supported by studies of the radiation pattern of a heavy quark \cite{Hatta,Chernicoff3}. See \cite{Chernicoff4} for a review of all these topics.}. \\

 We conclude this section with some more remarks on the zero temperature dissipation term: \\
 \begin{itemize}
 \item It is {\em finite} and cannot be renormalized away in the boundary theory by Hermitian counter terms. \\
 \item
There has been some discussion in the literature on zero temperature dissipation \cite{Pitaevskii,Mohanty,Bonart,Cherny,Walmsley,Kamenev,KMN}. \cite{KMN} advocate renormalizing this term away by subtracting the contribution from pure AdS which corresponds to a vacuum. While this is certainly a valid option, we do not feel compelled to do this, because as we have seen in this 1+1 dimensional system there is no violation of any physical principle such as Lorentz invariance.  Also 
as the calculations of radiation show, there is a compelling physical reason to expect that it should be there.\\
\item We take the view point that $\mu$ is finite because it is to be interpreted as an RG scale.  Also as mentioned earlier, as $\mu \to \infty$ the particle becomes infinitely heavy. Otherwise there is nothing pathological in our calculation even if $\mu$ is infinite. \\

\item There are some 1+1 condensed matter systems \cite{Pitaevskii,Mohanty,Bonart,Cherny,Walmsley,Kamenev}\footnote{The authors would like to thank G. Baskaran for pointing out these references.} which exhibit such dissipation (or decoherence) at absolute zero due to zero-point fluctuations. 
\end{itemize}

\section{Schwinger-Keldysh Propagators from Holography : A Review} \label{sec:SK}
The first two subsections of this section are basically review of how to get Schwinger-Keldysh propagators in the boundary field theories using extended Kruskal structure of the black hole. This is written in terms of the retarded Green functions. Thus combining this with the results of Section \ref{sec:TS}, we immediately obtain the exact Schwinger-Keldysh Green functions for our system.   
\subsection{Kruskal/Keldysh correspondence}
\begin{figure}
\begin{center}
\begin{tikzpicture}
\draw [blue] (-4,0)--(-4,-7);
\draw [blue] (4,-7)--(4,0);
\draw (-4,-7)--(0,-3.5)node[pos=.5,sloped,above] {$U=0$};
\draw (0,-3.5)--(4,0) node[pos=.5,sloped,above] {$U=0$};
\draw (4,-7)--(0,-3.5)node[pos=.5,sloped,above] {$V=0$};
\draw (0,-3.5)--(-4,0) node[pos=.5,sloped,above] {$V=0$};
\draw [red,snake=snake,
segment amplitude=1mm,
segment length=4mm,
line after snake=0mm](-4,0) -- (4,0);
\draw [red,snake=snake,
segment amplitude=1mm,
segment length=4mm,
line after snake=0mm](-4,-7) -- (4,-7);
\draw[dashed,<->] (0,1)--(0,-8);
\draw[dashed,<->] (-5,-3.5)--(5,-3.5);
\draw [red](2,-3.5) node{\Large{R}};
\draw [red](-2,-3.5) node{\Large{L}};
\draw (0.5,.8) node{t$_K$};
\draw (4.9,-3.9) node{\large{x}$_K$};
\draw (3,-3.2) node{$V>0$};
\draw (3,-3.8) node{$U<0$};
\draw (-3,-3.2) node{$U>0$};
\draw (-3,-3.8) node{$V<0$};

\end{tikzpicture}
\caption{\emph{AdS space in Kruskal coordinates}} \label{fig:Kruskal}
\end{center}
\end{figure}
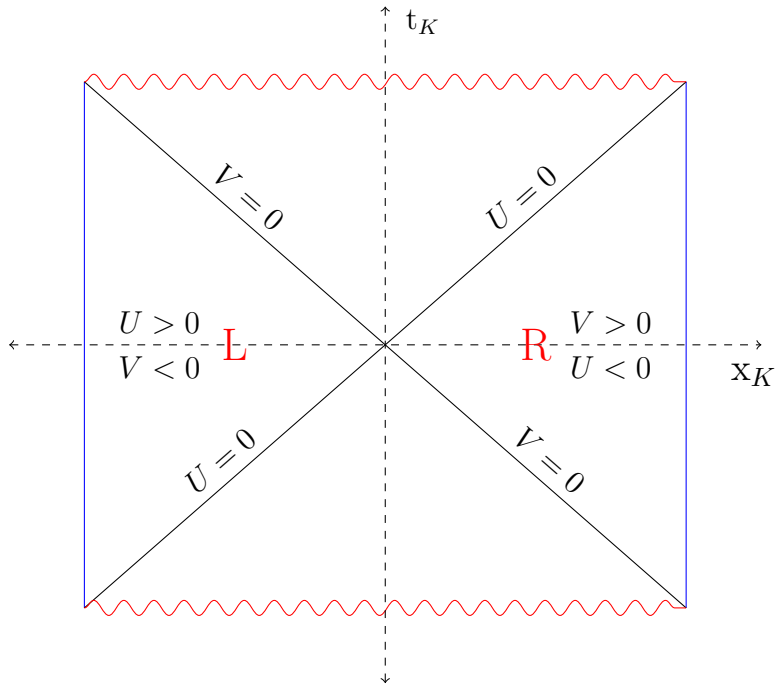

Herzog and Son \cite {HS} derived Schwinger-Keldysh propagators from bulk calculation in AdS$_5$-Schwarzschild metric. They analytically continued \cite{BD,Unruh} the modes of the scalar field from I to II (see figure \ref{fig:Kruskal}). During this procedure only the modes near the horizon are crucial. It is very straight forward to see that their prescription goes through for AdS$_3$-BTZ background too, as modes near the horizon behave identically. The same method is also applicable to our system with string where modes are functions of frequency ($\omega$) only. Their derivation involved symmetric contour  i.e, $\sigma = \beta/2~$. As we want to express our result in $ra$ formalism we will fix $ \sigma=0 $ as before.  We just sketch the generic four step AdS/CFT procedure . 

1. The EOM for the fluctuating string is solved subjected to the boundary conditions 
\begin{align}\label{eq:BP:KK1}
&\lim_{r\to r_m}x(\omega , r_1)=x_1^0(\omega) \\ \label{eq:BP:KK2}
&\lim_{r\to r_m}x(\omega , r_2)=x_2^0(\omega)
\end{align}
Here r$_1$ , r$_2$ are the radial coordinates in L and R regions respectively. Now the general solutions in L and R are 
\begin{align}\label{eq:BP:KK3}
&x(\omega , r_1)=a(\omega)f_{\omega}(r_1)+ b(\omega)f_{\omega}^*(r_1) \\ \label{eq:BP:KK4}
&x(\omega , r_2)=c(\omega)f_{\omega}(r_2)+ d(\omega)f_{\omega}^*(r_2)
\end{align}

2. We have four undetermined coefficients in \eqref{eq:BP:KK3} and \eqref{eq:BP:KK4} but have only two boundary conditions namely \eqref{eq:BP:KK1}and \eqref{eq:BP:KK2}. So to specify the solution uniquely we need other two constraints . Imposing horizon boundary conditions we can eliminate two coefficients namely $c(\omega)$ and $d(\omega)$. Near the horizon the ingoing and outgoing modes in the R region behave as
\begin{align}\label{eq:BP:KK5}
&e^{-i\omega t}f_{\omega}(r_1) \sim e^{-i\frac{\omega}{2\pi T}log(V)} \\ \label{eq:BP:KK5.1}
&e^{-i\omega t}f_{\omega}^*(r_1) \sim e^{+i\frac{\omega}{2\pi T}log(-U)}
\end{align}
Now following \cite{HS} we will analytically continue the solution from R (U $<$ 0, V $>$ 0 ) to L( U $>$ 0, V $<$ 0 ) region such that the solution is analytic in lower V plane and upper U plane\footnote{This choice is motivated by the fact that in field theory Feynman Green function contains +ve energy modes for $t \to \infty$ and -ve energy modes for $t \to -\infty$. And the Green function ($ G_{11} $) for the field theory ``living'' on the boundary of the R-region should be time ordered one like usual Feynman Green function,$G_F$ in Minkowski space. }. 
\begin{align}\label{eq:BP:KK6}
&f_{\omega}(r_1) \to e^{-\omega/2T}f_{\omega}(r_2) \\ \label{eq:BP:KK6.1}
&f_{\omega}^*(r_1) \to e^{+\omega/2T}f_{\omega}^*(r_2)
\end{align}
Therefore the solution when analytically continued to L region becomes 
\begin{align}\label{eq:BP:KK7}
x(\omega , r_2)=a(\omega)e^{-\omega/2T}f_{\omega}(r_2)+ b(\omega)e^{+\omega/2T}f_{\omega}^*(r_2)
\end{align}
But as mentioned above this is a special case where the contour is symmetric i.e, $\sigma= \beta/2$. One can generalize this result by starting with $V \to |V|e^{-i\theta}$ and $-U \to |U| e^{-i(2\pi - \theta)}$ and defining $\sigma \equiv \frac{\theta}{2 \pi T}$ , then continuing analytically to get
\begin{align}\label{eq:BP:KK8}
&f_{\omega}(r_1) \to e^{-\omega \sigma} f_{\omega}(r_2) = f_{\omega}(r_2) \\ \label{eq:BP:KK8.1}
&f_{\omega}^*(r_1) \to e^{+\omega/T}e^{-\omega \sigma}f_{\omega}^*(r_2) = e^{+\omega/T}f_{\omega}^*(r_2)
\end{align}
here we have taken $\sigma=0$ as usual.
So, the solution in L region reduces to
\begin{align}\label{eq:BP:KK9}
x(\omega , r_2)=a(\omega)f_{\omega}(r_2)+ b(\omega)e^{+\omega/T}f_{\omega}^*(r_2)
\end{align}

Now imposing the boundary conditions \eqref{eq:BP:KK1},\eqref{eq:BP:KK2} into \eqref{eq:BP:KK3} and \eqref{eq:BP:KK9} we can solve for $a(\omega)$ and $b(\omega)$ 
\begin{align}\label{eq:BP:KK10}
&a(\omega)= x_1^0(\omega)\{1+n_B(\omega)\}-x_2^0(\omega)n_B(\omega)\\ \label{eq:BP:KK10.1}
&b(\omega)= x_2^0(\omega)n_B(\omega)-x_1^0(\omega)n_B(\omega)
\end{align}
Now we have \emph{the} solution fully specified by R and L region solutions  \eqref{eq:BP:KK3} and \eqref{eq:BP:KK9} satisfying necessary boundary conditions at the boundary and the horizon .\\

3. The next step is to plug this solution into the boundary action 
\begin{align}\label{eq:BP:KK10}
S_{\text{b'dy}}=-\frac{T_0(r_m)}{2}\int_{r_1}\frac{\mathrm d{\omega}}{2\pi}~x_1(-\omega,r_1)~\partial_rx_1(\omega,r_1)+\frac{T_0(r_m)}{2}\int_{r_2}\frac{\mathrm d{\omega}}{2\pi}~x_2(-\omega,r_2)~\partial_rx_2(\omega,r_2)
\end{align}
to get 
\begin{align}\label{eq:BP:KK11}
iS_{\text{b'dy}}=-\frac{1}{2}\int&\frac{\mathrm d{\omega}}{2\pi}~x_1^0(-\omega)\left[i\text{Re}~G_R^0-(1+2n_B)~\text{Im}~G_R^0\right]~x_1^0(\omega) \nonumber \\ \nonumber
&+x_2^0(-\omega)\left[-i\text{Re}~G_R^0-(1+2n_B)~\text{Im}~G_R^0\right]~x_2^0(\omega) \\ \nonumber
&-x_1^0(-\omega)~\left[-2n_B~ \text{Im}~G_R^0\right]~x_2^0(\omega) \\ 
&-x_2^0(-\omega)~\left[-2(1+n_B)~\text{Im}~G_R^0\right]~x_1^0(\omega)
\end{align}

Here retarded Green function is defined as 
\begin{align}\label{eq:BP:KK12}
G_R^0(\omega)\equiv ~T_0(r)~\frac{f_{-\omega}(r)~\partial_rf_\omega}{|f_{\omega}(r)|^2}\bigg|_{r=r_m}
\end{align}
(as $r \to \infty $ , $|f_{\omega}(r)|^2 \to 1$ . So the numerator is already normalized if the probe brane is very close to the boundary.) \\

4. The last step is to take functional derivative with respect to $x_1^0$ and/or $x_2^0$ which are acting like two source terms for the boundary field theory to get the Schwinger-Keldysh propagators
\begin{align}\label{eq:BP:KK13}
G_{\text{ab}}=\begin{bmatrix} i\text{Re}~G_R^0-(1+2n_B)~\text{Im}~G_R^0& -2n_B ~\text{Im}~G_R^0\\ -2(1+n_B)~\text{Im}~G_R^0 & -i\text{Re}~G_R^0-(1+2n_B)~\text{Im}~G_R^0\end{bmatrix} 
\end{align}

$G_{\text{ab}}$ is exactly known from the expressions of $G_R^0$ in \eqref{eq:TL:GF2} and \eqref{eq:TL:GF3.1}. \\

Now we want to express our result in $ra$ formalism. So we need to covert $x_1,x_2$ into $x_r,x_a$. Then the relations between bulk and the boundary fields reduce to 
\begin{align}\label{eq:BP:KK14}
&x_a(\omega,r)=f_{\omega}^*(r)x_a^0(\omega) \\ \label{eq:BP:KK14.1}
&x_r(\omega,r)=f_{\omega}(r)x_r^0(\omega)+i(1+2n_B)\text{Im}f_\omega(r)x_a^0(\omega)
\end{align}
And the boundary action in this set up becomes 
\begin{align}\label{eq:BP:KK15}
S_{\text{b'dy}}=\displaystyle -~\frac{T_0(r_m)}{2}\int_{r_m}\frac{\mathrm d{\omega}}{2\pi}x_a(-\omega,r)~\partial_rx_r(\omega,r)-~\frac{T_0(r_m)}{2}\int_{r_m}\frac{\mathrm d{\omega}}{2\pi}x_r(-\omega,r)~\partial_rx_a(\omega,r)
\end{align}
Plugging  \eqref{eq:BP:KK14} and \eqref{eq:BP:KK14.1} into the boundary action as before we end up getting
\begin{align}\label{eq:BP:KK16}
iS_{\text{b'dy}}=-i\int \frac{\mathrm d{\omega}}{2\pi}x_a^0(-\omega)~[G_R^0(\omega)]~x_r^0(\omega)-\frac{1}{2}\int\frac{\mathrm d{\omega}}{2\pi}x_a^0(-\omega)~[iG_{\text{sym}}(\omega)]~x_a^0(\omega)
\end{align}

\subsection{Boundary stochastic motion}

Now here we want to have the boundary stochastic motion of the string. The partition function for the string can be written as 
\begin{align}\label{eq:BP:BSM1}
Z= \int [\mathfrak{D}x_1 \mathcal{D}x_1^0 ]~ [\mathfrak{D}x_2 \mathcal{D}x_2^0]~ e^{iS_1-iS_2}
\end{align}
where $\mathcal{D}x_1^0$ is a measure for temporal path of the string end point and $\mathfrak{D}x_1$ is a measure for the bulk path integral for the body of the string in R-region of the full Kruskal plane. Similarly $\mathcal{D}x_1^0$  and $\mathfrak{D}x_1$ are defined in L-region. 
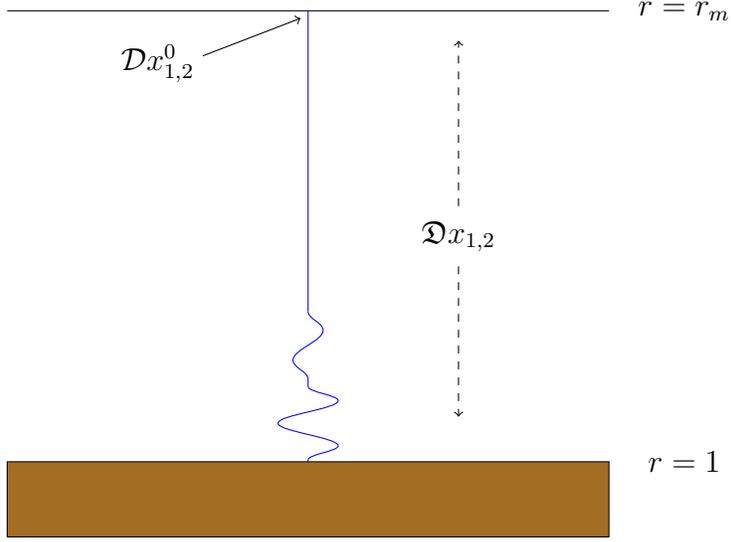
\begin{figure}
\begin{center}
\begin{tikzpicture}
\draw (-4,0) to (4,0);
\draw[fill=brown1] (-4,-7) rectangle (4,-6);
\draw [blue] (0,0) -- (0,-4);
\draw [blue,snake=snake,
segment amplitude=2mm,
segment length=8mm,
line after snake=0mm](0,-4) -- (0,-5);
\draw [blue,snake=snake,
segment amplitude=4mm,
segment length=6mm,
line after snake=0mm](0,-5) -- (0,-6);

\draw (-2,-0.7) node{$\mathcal{D}x^0_{1,2} $};
\draw [->] (-1.4,-0.6)--(-0.1,-0.1);

\draw[dashed,->] (2,-2.6)--(2,-0.4);
\draw[dashed,->] (2,-3.4)--(2,-5.4);

\draw (2,-3.0) node{$\mathfrak{D}x_{1,2} $};

\draw (5,0) node{$r=r_m$};
\draw (5,-6) node{$r=1$};
\end{tikzpicture} 
\caption{\emph{Visualizing the boundary stochastic motion of the heavy particle by integrating out all string degrees of freedom.}} \label{fig:Bulk1}
\end{center}
\end{figure}
\begin{align} \label{eq:BP:BSM2}
[\mathcal{D}x_1^0]= \prod_t \text{d} x_1^0(t) , \hspace{2cm} [\mathfrak{D}x_1]= \prod_{t,r} \text{d} x_1(t,r)
\end{align}

To obtain the effective action of the string end points we will integrate out all string coordinates inside the bulk.
If we do this path integral (over the terms contained in the bracket)
\begin{align}\label{eq:BP:BSM3}
Z&=\displaystyle \int [\mathcal{D}x_1^0 ]~ [\mathcal{D}x_2^0]~ \underbrace{[\mathfrak{D}x_1]~ [\mathfrak{D}x_2]~   e^{iS_1-iS_2}} \nonumber \\ 
&\equiv \int  [\mathcal{D}x_1^0 ] ~[\mathcal{D}x_2^0] e^{i S_{\text{eff}}^0}
\end{align}
We have absorbed the field independent determinant in the normalization of the path integral.  Now will use the results from previous section where we have already calculated the boundary actions \eqref{eq:BP:KK10} and \eqref{eq:BP:KK15}.
As there is no ``boundary'' at the horizon there will be only  two boundary terms from the two boundaries of the Kruskal plane. 
\begin{align}\label{eq:BP:BSM4}
S_{\text{eff}}^0= -~\frac{T_0(r_m)}{2}\int_{r_1}\frac{\mathrm d{\omega}}{2\pi}x_1(-\omega,r_1)\partial_rx_1(\omega,r_1)+\frac{T_0(r_m)}{2}\int_{r_2}\frac{\mathrm d{\omega}}{2\pi}x_2(-\omega,r_2)\partial_rx_2(\omega,r_2)
\end{align}

Now we can easily write down the partition function for the string endpoints in $ra$ formalism from \eqref{eq:BP:KK16}
\begin{align}\label{eq:BP:BSM5}
Z=\int  [\mathcal{D}x_r^0 ] ~[\mathcal{D}x_a^0] ~e^{i S_{\text{eff}}^0}
\end{align} 
\begin{align}\label{eq:BP:BSM6}
iS_{\text{eff}}^0=-i\int \frac{\mathrm d{\omega}}{2\pi}x_a^0(-\omega)[G_R^0(\omega)]x_r^0(\omega)-\frac{1}{2}\int\frac{\mathrm d{\omega}}{2\pi}x_a^0(-\omega)[~iG_{\text{sym}}(\omega)]x_a^0(\omega)
\end{align}

Notice that the effective partition function of the string end points \eqref{eq:BP:BSM5} is exactly similar to the Fourier space path integral \eqref{eq:LE:13}. Therefore we can perform the same procedure of introducing a ``noise'', $\xi$ to obtain the following equations of motion obeyed by the string end points 
\begin{align}\label{eq:BP:BSM7}
\left[- M_Q^0 \omega^2 + G_R(\omega)\right] x_r(\omega)= \xi(\omega) \hspace{2cm} \langle\xi(-\omega)\xi(\omega) \rangle = -(1+2n_B)\text{Im}G_R(\omega)
\end{align}
Here we have used the facts that
\begin{align*}
&G_R^0(\omega)= - M_Q^0 \omega^2 + G_R(\omega),\\
&i G_{\text{sym}}(\omega)=-(1+2n_B)\text{Im}G_R(\omega)
\end{align*}

\section{Effective Action at General $r$ : Brownian Motion on Stretched Horizon} \label{sec:EA}
Since we have an exact solution one can hope to generalize the membrane paradigm by locating the membrane at {\em arbitrary}
$r$ ($r_0$, say). This would be in the spirit of a holographic renormalization group (RG) \cite{Strominger,FLR} approach to the problem. This would then
justify the statement made in Section \ref{sec:TS} that $\mu$ can be interpreted as an RG scale. This is done in this section.
We begin with a review of some basic ideas in holographic RG following \cite{FLR}.
\subsection{Holographic Renormalization Group}
\label{sec:HRG}

A version of the holographic RG that is useful here was discussed in \cite{FLR} and is reviewed in this section. The main idea is to start with an action, which is the original bulk action supplemented by a boundary action at $r=r_0$, that takes into account the effect of integrating out of the bulk region $r>r_0$.\footnote{In this subsection for compatibility with \cite{FLR} we use $r$ for the dimensional variable. This would be $\bar r$ in other sections.} This region $r>r_0$ in the bulk represents
the high energy region of the boundary field theory.  The so called ``alternative quantization'' \cite{Klebanov-Witten} where the boundary value of the bulk field $\phi$ is interpreted as the expectation value of a boundary single trace operator rather than as a source for the boundary operator comes in handy in explaining the approach.  The boundary action obtained this way can also be interpreted as
the generating functional for a different boundary theory that is obtained by the so called  ``standard quantization''. Furthermore there is an RG flow from the first boundary action perturbed by a relevant deformation involving double trace operators to the second boundary action.

Thus we begin with  
\begin{align}  \label{phi-action}
S= \int_0^{r_0} \mathrm d{r} ~\mathrm d^D{x}~ \sqrt {-g}\left[-~{1 \over 2} ~\p_M\phi ~\p_N \phi ~g^{MN} -V(\phi)\right] + S_B[\phi,r_0]   
\end{align}

In the \cite{FLR} $D$ is the space time dimension of the boundary theory and $\phi$ fills all of $AdS$ bulk. However we can interpret
$D$ for our purposes as the dimension of a brane/string hanging down from the boundary into the center with the other end
going into the horizon of the black hole. Thus in our case $\phi(x)=x(r,t)$, $D=1$ and the action becomes
\begin{align} \label{x-action}
S= \int_0^{r_0} \mathrm d{r} ~\mathrm dt~ \sqrt {-g}\left[-~{1 \over 2} ~\left(\p_t x(r,t)~\p_t x(r,t) ~g^{tt}+\p_r x(r,t)~\p_r x(r,t) ~g^{rr}\right) -V(x)\right] + S_B[x,r_0]   
\end{align}
This can be compared with \eqref{Action} and we see that it is exactly the same with $V(x) =m$, which does not contribute to the equations of motion, and so can be ignored in this discussion.

For our purposes, since we are only interested in the two point function, we can think of the boundary action as
\be   \label{eq:SB}
S_B[\phi,r_0] = {1\over 2} \int _{r=r_0} \mathrm d^D{k}~ \phi(k)G_R(k,r_0) \phi(-k)
\ee 
Specializing to our case this becomes:
\be   \label{eq:SBx}
S_B[x,r_0] = {1\over 2} \int _{r=r_0} \mathrm d\omega~ x(\omega)G_R(\omega,r_0) x(-\omega)
\ee 
 
The parameters of the boundary field theory action are collected here in $G_R(k,r_0)$ and their dependence on the RG scale
$r_0$ is indicated. 
When we vary $\phi$ we get the usual bulk equation and also a (boundary) condition at the boundary $r=r_0$.   This depends on the boundary action and is:
\be	\label{eq:GR}
G_R(k,r_0)= -\sqrt{-g} g^{rr}{\p_r\phi_c(r_0) \over \phi_c(r_0)}
\ee
Fixing the solution to the equation of motion, a second order differential equation in $r$, requires specifying $\phi(r_0)$ and $\p_r \phi(r_0)$. 
If we specify $\phi(r_0)$ and $\p_r \phi(r_0)$, $G_R$ is fixed by this boundary condition. In the alternative quantization $G_R$ is
the coefficient of the quadratic term in the effective action of the boundary theory. On the other hand if we interpret $S_B(\phi)$
as the generating functional for the boundary theory, $G_R(k, r_0)$ is the Green function of the boundary theory.  This is the interpretation that is relevant for us. The Green functions in the two cases are inverses of each other.

One important point is that if the bulk equation of motion is linear, therefore scaling $\phi(r_0)$ by a number just scales the solution everywhere by the same number. Hence $G_R$ is not affected. But this would not be true in a non linear bulk theory. In Sec \ref{sec:TS} we have a linear approximation to the equation for the string fluctuation. Thus there is no loss of generality in setting $\phi(r_0)=1$. 

In this approach one can write down an RG,  \cite{FLR}, that says the total action (evaluated on the solution) cannot depend on $r_0$. As also shown in \cite{FLR} the parameters of the boundary action must vary such that the the classical solution is reproduced. Thus solving the RG gives the classical solution. The converse is also true. It is easy to see \cite{FLR} that if we use the exact classical solution in the action, the RG becomes an identity, because it becomes equivalent to imposing \eqref{eq:GR}.

The above formalism can be applied to our case where we use \eqref{x-action} and \eqref{eq:SBx}. 
\be	\label{eq:GRx}
G_R(\omega,r_0)= -\sqrt{-g} g^{rr}{\p_r x_c(r_0) \over x_c(r_0)}
\ee
with $\sqrt{-g} g^{rr}=T_0$. This is the same as \eqref{RGF} except that we have not assumed any normalization for the $x_c(r_0)$. 

As we change $r_0$ to $r_0'$, RG demands that one has to change the boundary condition on $x$ and the boundary action so that physical quantities are fixed. In our case since we know the exact solution, we know the boundary condition at $r_0'$:  $x(r_0') =x_c (r_0')$ where $x_c$ is the exact classical solution, which has an earlier prescribed boundary value at $r_0$. We also know the new boundary action. It is given by \eqref{eq:SB} where $G_R(r_0')$ is given by \eqref{eq:GR}, where the RHS is evaluated at $r_0'$. 
(Actually for the situation in Section \ref{sec:TS},  the equation for $x$ is linear, and as mentioned above we can just continue to use $x(r_0')=1$.)
 The functional form of the Green function does not change - except that all explicit $r_0$'s are replaced by $r_0'$'s. Thus the parameter $\mu = {r_0\over l_s^2}$ used in Section \ref{sec:TS} can be understood as a renormalization scale.

 Thus $G_R^0$ in our case is the correlation function
for the random force  i.e. we interpret the action involving $x$ as the generating functional for the boundary theory of the random force acting on the quark. 

 \eqref{eq:TL:GF1.1} has a diffusive pole at $ -i \mu\over \sqrt{\lambda}$. This gives an exponential decay time scale for the
 random force acting on the quark. Being a mass scale it is appropriately proportional  to $\mu$ the RG scale.  From the point of view
 of the action for $x$ (which is the coordinate of the quark in addition to being the source for $\xi$), this is a non local quadratic term
 and cannot be renormalized away by adding local counter terms. For the effective action for $\xi$ the random force acting on the quark, which involves the inverse Green function, this is a zero rather than a pole. However being imaginary, it cannot be renormalized away by a hermitian counter term in the bare action, and furthermore the powers of $\omega$ in the denominator would make the counterterm non local.
 This leads us to conclude that this pole  represents a  physical effect in the low energy dynamics of the quark.

\subsection{Placing the Membrane at Arbitrary $r$}

In the previous section we have integrated out all modes of the string to obtain the effective action for the string end points and we end up getting a Langevin equation on the boundary. Here our aim is to obtain such an effective action on a spatial slice at a general value of $r$ ($r_0$, say). This requires determining the solution to the EOM \eqref{eq:TS:EOM} exactly which we have already obtained \eqref{eq:TL:mode}. Then we will choose $r_0$ very close to $r_h$ to get a Langevin equation on that stretched horizon (a hypothetical membrane which we consider to be very close to the horizon of the black hole).

\par
\begin{figure}
\begin{center}
\begin{tikzpicture}

\draw (-4,0) to (4,0);
\draw[fill=brown1] (-4,-7) rectangle (4,-6);
\draw[fill=white,red!6] (-4,-6) rectangle (4,-4.0);

\draw [dashed][blue](-4,-5.6) -- (4,-5.6);
\draw [dashed][blue](-4,-5.605) -- (4,-5.605);
\draw [dashed][blue](-4,-5.61) -- (4,-5.61);

\draw [dashed][red](-4,-4.0) -- (4,-4.0);
\draw [blue] (0,0) -- (0,-4);
\draw [blue,snake=snake,
segment amplitude=2mm,
segment length=8mm,
line after snake=0mm](0,-4) -- (0,-5);
\draw [blue,snake=snake,
segment amplitude=4mm,
segment length=6mm,
line after snake=0mm](0,-5) -- (0,-6);

\draw (-2,-0.7) node{$\mathcal{D}x^0_{1,2} $};

\draw [->] (-1.5,-0.7)--(-0.1,-0.1);

\draw (-2,-5.0) node{$\mathcal{D} x^{r_0}_{1,2} $};

\draw[->](-1.45,-4.8)--(-.22,-4.2);

\draw[dashed,->] (2,-1.6)--(2,-0.1);
\draw[dashed,->] (2,-2.4)--(2,-3.8);

\draw (2,-2.0) node{$\mathfrak{D}x^>_{1,2} $};

\draw (1.6,-5.0) node{$\mathfrak{D}x^{<}_{1,2} $};


\draw[dashed,->] (1.6,-4.6)--(1.6,-4.1);
\draw[dashed,->] (1.6,-5.3)--(1.6,-5.9);

\draw (5,0) node{$r=r_m$};
\draw (5,-6) node{$r=1$};
\draw (5.3,-5.6) node{$r=1+\epsilon$};
\draw (5.1,-4.0) node{$r=r_0$};
\end{tikzpicture}
\caption{\emph{Integrating out the string degrees of freedom inside a hypothetical ``membrane'' and push it  very close to the  horizon (i.e, stretched horizon) to obtain a Langevin equation which is overdamped.}} \label{fig:Bulk2}
\end{center}
\end{figure}
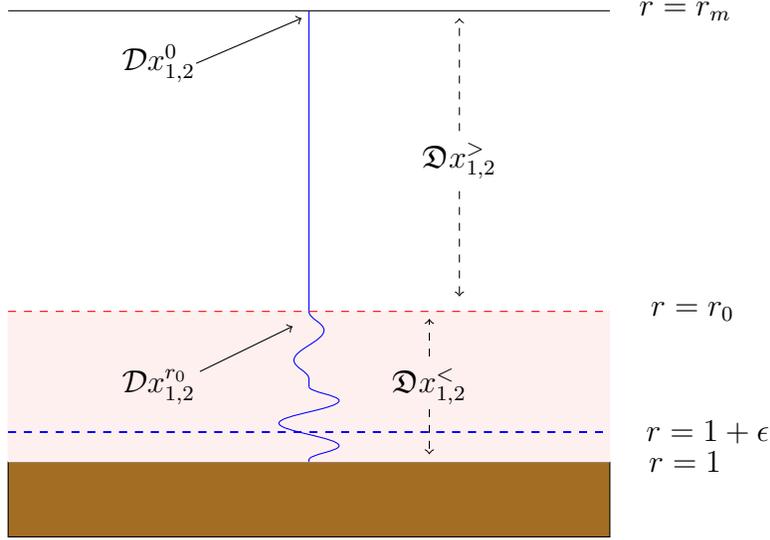

 For this purpose again we write the partition function of string in several parts
\begin{align} \label{eq:BM:2}
Z&=\displaystyle \int  [\mathcal{D}x_1^0 \mathfrak{D}x^>_1 \mathcal{D}x_1^{r_0}] ~[\mathcal{D}x_2^0 \mathfrak{D}x^>_2 \mathcal{D}x_2^{r_0}]~[\mathfrak{D}x^{<}_1\mathfrak{D}x^{<}_2] ~ e^{iS^>_1-iS^>_2}~ e^{iS^{<}_1-iS^{<}_2}  
\end{align}
As before $\mathcal{D}x_1^0$ is a measure for temporal path of the string end point and $\mathfrak{D}x^>_1$ and $\mathfrak{D}x^<_1$ are the measures for the bulk path integral for the body of the string outside and inside of the spatial slice in R-region  and $\mathcal{D}x_1^{r_0}$ denotes the temporal path integral for the string end point on the spatial slice (see fig. \ref{fig:Bulk2}). Whereas $S^>_1$ is the action outside the spatial slice and $S^<_1$ is the action inside the spatial slice. Same is true for L region. This time integrating out the region of the string inside $r=r_0$. 

\begin{align} \label{eq:BM:3}
Z&=\displaystyle \int [\mathcal{D}x_1^0 \mathfrak{D}x^>_1\mathcal{D}x_1^{r_0}] ~[\mathcal{D}x_2^0 \mathfrak{D}x^>_2 \mathcal{D}x_2^{r_0}]~ e^{iS^>_1-iS^>_2}\underbrace{[\mathfrak{D}x^{<}_1] ~[\mathfrak{D}x^{<}_2] ~e^{iS^{<}_1-iS^{<}_2}}  \\ \label{eq:BM:3.1}
&=\displaystyle \int [\mathcal{D}x_1^0 \mathfrak{D}x^>_1\mathcal{D}x_1^{r_0}] ~[\mathcal{D}x_2^0 \mathfrak{D}x^>_2 \mathcal{D}x_2^{r_0}]~ e^{iS^>_1-iS^>_2}~e^{i S_{\text{eff}}^{r_0}}
\end{align}

Where $S^{r_0}_\text{eff}$ = The boundary action which passes through $x_1^{r_0}(\omega)$ and $x_2^{r_0}(\omega)$. \\ 
Notice here the stretched horizon at $r=r_0$ is a boundary. And the older boundary conditions are now applicable at $r=r_0$. So the the bulk fields, $x^{<}_{1,2}(r,\omega)$ and the boundary fields, $x^{r_0}_{1,2}(\omega)$ are related by
\begin{align} \label{eq:BM:4}
&x^{<}_{1}(r_1=r_1^0,\omega)=x^{r_0}_{1}(\omega) \\\label{eq:BM:4.1}
&x^{<}_{2}(r_2=r_2^0,\omega)=x^{r_0}_{2}(\omega)
\end{align}
And if we use the $ra$ basis then the boundary conditions reduce to
\begin{align}\label{eq:BM:5}
&x^<_a(\omega,r)=f_{\omega}^*(r)x_a^{r_0}(\omega) \\ \label{eq:BM:5.1}
&x^<_r(\omega,r)=f_{\omega}(r)x_r^{r_0}(\omega)+i(1+2n_B)\text{Im}f_\omega(r)x_a^{r_0}(\omega)
\end{align}
Going through the same calculation as before and using the fact there is no ``boundary'' at the horizon we end up getting the membrane effective action
\begin{align} \label{eq:BM:6}
S^{r_0}_\text{eff}= -\frac{T_0(r_0)}{2}\int_{r^0_1}\frac{\mathrm d{\omega}}{2\pi}x^<_1(-\omega,r)\partial_r x^<_1(\omega,r)+\frac{T_0(r_0)}{2}\int_{r^0_2}\frac{\mathrm d{\omega}}{2\pi}x^<_2(-\omega,r)\partial_r x^<_2(\omega,r)
\end{align}
Now if we use  the $ra$-coordinates then using \eqref{eq:BM:5} and \eqref{eq:BM:5.1} we will have 
\begin{align}\label{eq:BM:7}
iS_{\text{eff}}=-i\int \frac{\mathrm d{\omega}}{2\pi}x_a^{r_0}(-\omega)[G_R^{r_0}(\omega)]x_r^{r_0}(\omega)-\frac{1}{2}\int\frac{\mathrm d{\omega}}{2\pi}x_a^{r_0}(-\omega)[~i G^{r_0}_{\text{sym}}(\omega)]x_a^{r_0}(\omega)
\end{align}
\par The retarded propagator is defined such that it is normalized at the spatial slice. Now the expression for retarded force-force correlator can be written down for any fixed value of $r_0$. Using the value of $T_0(r)$ from \eqref{Tension} and substituting the expression for $f_\omega(r)$ from \eqref{eq:TL:mode}
\begin{align} 
G_R^{r_0}(\omega)&\equiv T_0(r)~\frac{f_{-\omega}(r)\partial_rf_\omega}{|f_{\omega}(r)|^2}\bigg|_{r=r_0}  \nonumber \\ \label{eq:BM:8.1}
&= -~ \frac{\sqrt{\lambda}\pi^2 T^3}{2}~ \frac{ r_0 \mathfrak{w} (r_0 \mathfrak{w} +i)}{(r_0 -i \mathfrak{w} )}  \\\label{eq:BM:8.2}
&= -~\mu_0 \omega ~{(i \sqrt \lambda \pi ^2 T^2 + \mu_0 \omega )\over 2\pi  ( \mu_0 - i \sqrt \lambda \omega )} 
\end{align}

\eqref{eq:BM:8.1} and \eqref{eq:BM:8.2} are exact expressions for the retarded propagator on the probe brane which is placed at $r=r_0$ and or equivalently when the field theory is probed at the energy scale $\mu_0=\frac{\bar{r}_0}{l_s^2}$ . It trivially reduces to the boundary propagator $G_R^0(\omega)$ as in \eqref{eq:TL:GF1a} when $r_0 \to r_m$.\\

The other point we want to emphasis here is that the retarded Green function \eqref{eq:TL:GF3.1} which is derived using holography 
incorporates the ``softening of delta function'' to avoid the contradiction described in section \ref{sec:intro}. 
\begin{align} \label{delta2}
\displaystyle \lim _{t \to t_0} \int_{t_0}^t \mathrm d{t^\prime}~\gamma(t^\prime)&=\displaystyle \lim _{t\rightarrow t_0} \int_{t_0}^t \mathrm d{t^\prime}\int_{-\infty}^{\infty} \mathrm d{\omega}~ e^{-i\omega t'}\gamma(\omega) \\
&=- \displaystyle \lim _{t\rightarrow t_0} \int_{t_0}^t \mathrm d{t^\prime}\int_{-\infty}^{\infty} \mathrm d{\omega}~ e^{-i\omega t'} ~ {\mu \omega \over 2\pi}~{(\omega^2 + \pi ^2 T^2)\over (\omega + i {\mu\over \sqrt \lambda})}\end{align}

One can perform the contour integral for $\omega$ to pick up the residue at $\omega=~-i {\mu\over \sqrt \lambda}$ . So the corresponding integral 
\begin{align} \label{delta3}
\displaystyle \lim _{t \to t_0} \int_{t_0}^t \mathrm d{t^\prime} ~2\pi i~e^{-{\mu \over \sqrt{\lambda}} t'} ~ {\mu (-i {\mu\over \sqrt \lambda}) \over 2\pi}~{\left( \left(-i {\mu\over \sqrt \lambda}\right)^2 + \pi ^2 T^2 \right)} \to 0 
\end{align}
This shows that our Green function \eqref{eq:BM:8.2} is consistent with \eqref{delta}.

From \eqref{eq:BM:3.1} it is evident that after we integrate out the string coordinates inside $r=r_0$ we have partition functions of two haves of the Kruskal plane and which are coupled by the membrane effective action , $S^{r_0}_\text{eff}$ . Now will be shown (and already been mentioned) that a part ($G^{r_0}_{\text{sym}}$)of this ``coupling'' introduces thermal noise. It can be done following exactly same procedure of invoking a horizon noise for the second part of  $e^{iS^{r_0}_\text{eff}}$ in the partition function
\begin{align}\label{eq:BM:10}
&e^{-\frac{1}{2}\int \frac{\mathrm d{\omega}}{2\pi}~ x^{r_0}_a(-\omega)[i G^{r_0}_{\text{sym}}(\omega)]x^{r_0}_a(\omega)}= \int [\mathcal{D}\xi^{r_0}]~ e^{i \int x^{r_0}_a(-\omega) \xi^{r_0}(\omega)}~ e^{-\frac{1}{2}\int\frac{\xi^{r_0}(\omega)\xi^{r_0}(-\omega)}{i G^{r_0}_{\text{sym}}(\omega)}} \\
&\text{with ,~~~}  \langle\xi^{r_0}(-\omega)\xi^{r_0}(\omega) \rangle =~ i G^{r_0}_{\text{sym}}(\omega)=~-(1+2n_B)\text{Im}G^{r_0}_R(\omega) 
\end{align}
We have computed the contribution coming from the boundary action namely $S^{r_0}_\text{eff}$ . Now in order to calculate the partition function  \eqref{eq:BM:3.1} we need to look at the bulk contributions. In $ra$ basis this bulk action reduces to
\begin{align}\label{eq:BM:11}
iS^>_1-iS^>_2&= -i\int\frac{\mathrm d{\omega}}{2\pi}~dr\left[T_0(r)\partial_r x^>_a(-\omega,r) \partial_r x^>_r(\omega,r)-\frac{m\omega^2 x^>_r(\omega,r)x^>_a(-\omega,r)}{f}\right] \nonumber \\
&=-i\int\frac{\mathrm d{\omega}}{2\pi}~x^>_a(-\omega,r)~[T_0(r)\partial_r x^>_r(\omega,r)]\bigg|^{r=r_m}_{r=r_0} \nonumber \\
&\hspace{.5cm}-i\int\frac{\mathrm d{\omega}}{2\pi}~dr x^>_a(-\omega,r)\left[-\partial_r(T_0(r)\partial_r x^>_r(\omega,r))-\frac{m\omega^2 x^>_r(\omega,r)}{f}\right] \nonumber \\ 
\end{align}
From \eqref{eq:BM:7}, \eqref{eq:BM:10} and \eqref{eq:BM:11} we can finally write 
\begin{align}\label{eq:BM:12}
iS^>_1-iS^>_2+iS^{r_0}_{\text{eff}}= &-i\int_{r_m}\frac{\mathrm d{\omega}}{2\pi}~x^0_a(-\omega,r)~[T_0(r_m)\partial_r x^>_r(\omega,r)] \nonumber \\
&-i\int_{r_0}\frac{\mathrm d{\omega}}{2\pi}~x^{r_0}_a(-\omega,r)~[-T_0(r_0)\partial_r x^>_r(\omega,r)+G_R^{r_0}(\omega) x^{r_0}_r(\omega)-\xi^{r_0}(\omega)] \nonumber \\
&-i\int\frac{\mathrm d{\omega}}{2\pi}~dr ~x^>_a(-\omega,r)\left[-\partial_r\left(T_0(r)\partial_r x^>_r(\omega,r)\right)-\frac{m\omega^2 x^>_r(\omega,r)}{f}\right]
\end{align}

The path integral reduces to 
\begin{align}\label{eq:BM:13}
Z&= \displaystyle \int [\mathcal{D}x_r^0 \mathfrak{D}x^>_r\mathcal{D}x_r^{r_0}]~[\mathcal{D}\xi^{r_0}]~ e^{-\frac{1}{2}\int\frac{\xi^{r_0}(\omega)\xi^{r_0}(-\omega)}{-(1+2n_B)\text{Im}G^{r_0}_R(\omega)}} \underbrace{[\mathcal{D}x_a^0 \mathfrak{D}x^>_a \mathcal{D}x_a^{r_0}]~ e^{iS^>_1-iS^>_2+i S_{\text{eff}}^{r_0}}} \nonumber \\
&= \displaystyle \int [\mathcal{D}x_r^0 \mathfrak{D}x^>_r\mathcal{D}x_r^{r_0}] ~[\mathcal{D}\xi^{r_0}] ~e^{-\frac{1}{2}\int\frac{\xi^{r_0}(\omega)\xi^{r_0}(-\omega)}{-(1+2n_B)\text{Im}G^{r_0}_R(\omega)}}~\delta_\omega\left[-T_0(r_m)\partial_r x^>_r(\omega,r)\right]_{r=r_m}\nonumber \\ 
&\hspace{3cm}\delta_\omega\left[-\partial_r(T_0(r)\partial_r x^>_r(\omega,r))-\frac{m\omega^2 x^>_r(\omega,r)}{f}\right] \nonumber \\
&\hspace{3cm}\delta_\omega[-T_0(r_0)\partial_r x^>_r(\omega,r)+ G^{r_0}_R(\omega)x^{r_0}_r(\omega)-\xi^{r_0}(\omega)]_{r=r_0}
\end{align}

We have integrated over the terms inside the bracket viz, $[\mathcal{D}x_a^0], [\mathfrak{D}x^>_a]$ and $[\mathcal{D}x_a^{r_0}] $. This path integral \eqref{eq:BM:13} leads to three equations for boundary end point, the horizon end point and the body of the string. \\ \\
I. ~The boundary end point dynamics is governed by the deterministic equation which $~~~~~~$just tells us this end point is free 
\begin{align}\label{eq:BM:14}
T_0(r_m)\partial_r x^>_r(\omega,r)=0
\end{align}
II. ~The body of the string, as expected,  satisfies the equation of motion
\begin{align}\label{eq:BM:15}
\left[\partial_r\left(T_0(r)\partial_r x^>_r(\omega,r)\right)+\frac{m\omega^2 x^>_r(\omega,r)}{f}\right]=0
\end{align}
III. ~And the end point on the $r=r_0$ membrane obeys the stochastic equation of motion
\begin{align}\label{eq:BM:16}
T_0(r_0)\partial_r x^>_r(\omega,r) +\xi^{r_0}(\omega)= G^{r_0}_R(\omega)x^{r_0}_r(\omega) \\
\text{with,}~~~~ \langle \xi^{r_0}(-\omega) \xi^{r_0}(\omega) \rangle = -(1+2n_B) \text{Im} G^{r_0}_R(\omega)
\end{align}

This is the Langevin dynamics describing the endpoint of the string living on the membrane at $r=r_0$.

So far in this section everything was for arbitrary $r_0$ . Now for the sake of completeness we follow Son and Teaney \cite{Son-Teaney} again to re-derive the overdamped motion on the stretched horizon and also the boundary FD equation for AdS$_3$-BH which is identical to their (AdS$_5$-BH) calculation. For that purpose we want to move this membrane very close to the horizon i.e,  $r_0=1+\epsilon$ ; $\epsilon$ is very small (see fig. \ref{fig:Bulk2}). Putting this value of $r_0$ into \eqref {eq:BM:8.1} one obtains retarded Green function on the stretched horizon
\begin{align} \label{G_R^h}
G_R^h(\omega)&=-\displaystyle \lim_{\epsilon \to 0}\frac{\sqrt{\lambda}\pi^2 T^3}{2}\left[\frac{i(1+2\epsilon)(\mathfrak{w}+\mathfrak{w}^3)}{1+2\epsilon+\mathfrak{w}^2}+\frac{2\epsilon\mathfrak{w}^2}{1+2\epsilon+\mathfrak{w}^2}\right] \nonumber \\
&\sim -i\gamma \omega
\end{align}
Here we have assumed that frequency is very small i.e, $ \mathfrak{w}<< 1$. The other point to notice is the ``inertial term'' is suppressed by an extra factor of $\epsilon$ . And as $\epsilon \to 0$ the mass of the string end point on the stretched horizon , 
\begin{align}\label{eq:BM:9}
M^h_Q \equiv \frac{2\epsilon }{1+2\epsilon }\frac{\sqrt{\lambda}T^2}{2} \to 0   
\end{align}
( Expanding the real part of $G_R^h(\omega)$ in $\epsilon$ we get correction to mass , $\Delta M\sim \mathcal{O}(\epsilon^2)$. ) 
\par Therefore from \eqref{eq:BM:16} and \eqref{G_R^h} one obtains the Langevin equation on the stretched horizon
\begin{align}\label{eq:damped}
T_0(r_h&)\partial_r x^>_r(\omega,r) +\xi^{h}(\omega)= -i \omega \gamma x^{h}_r(\omega) \\ \label{eq:dampedFD}
\text{with,}~~~~~~~~~~~~~&\langle \xi^{h}(-\omega) \xi^{h}(\omega) \rangle = (1+2n_B)\gamma \omega
\end{align}
This is the overdamped motion of the horizon end point as discussed in \cite{Son-Teaney}, where the first term signifies the pulling of the end point by the string outside the horizon. Vanishing of M$^h_Q$ \eqref{eq:BM:9} is the reason behind getting an overdamped Langevin dynamics. \\

Our next task is to investigate how the fluctuations on this membrane is transmitted to the boundary through the string dynamics such that the boundary end point satisfies a Langevin equation \eqref{eq:BP:BSM7}. In other words, we wish to have a relationship between $\xi^{h}$ and $\xi^0$ and using this we want to show the fluctuation-dissipation for boundary fluctuations, $\xi^0$. To proceed let's consider the behavior of the solution near the AdS boundary
\begin{align}\label{eq:BM:17}
x(\omega , r)= x_0(\omega )f_\omega(r)+\xi^0(\omega )\left[\frac{\text{Im}f_\omega(r)}{-\text{Im}G_R(\omega)}\right]
\end{align}
where $f_\omega(r)$ is non-normalizable and $\text{Im}f_\omega(r)$ is a normalizable mode. $-\text{Im}G_R(\omega)$ is just a normalization such that $\xi^0(\omega )$ can be recognized as the boundary fluctuation. Now if substitute this \eqref{eq:BM:17} into the equation describing boundary dynamics \eqref{eq:BM:14} we obtain expected Brownian equation for the boundary end point
\begin{align}\label{eq:BM:18}
[-M^0_Q\omega^2 +G_R(\omega)]x_0(\omega )= \xi^0(\omega ) 
\end{align} \\
To get the fluctuation-dissipation relation for $\xi^0$ we use 
\begin{align}\label{eq:BM:19}
-i\omega \gamma = T_0(r_h)\frac{f_{-\omega}(r_h)\partial_rf_\omega(r_h)}{|f_{\omega}(r_h)|^2}
\end{align}
to re-write the equation at the stretched horizon dynamics \eqref{eq:damped} as 
\begin{align}\label{eq:BM:20}
\frac{\xi^0(\omega)}{-\text{Im}G_R(\omega)}T_0(r_h)\left[f_{\omega}(r_h)\partial_r \text{Im}f_{\omega}-\text{Im}f_{\omega}(r_h)\partial_r f_{\omega}(r_h)\right]+f_{\omega}(r_h)\xi^h(\omega)=0
\end{align} 
But the term in the square bracket is the Wronskian \eqref{Wronskian} of the two solutions and using $T_0(r)W(r)=+\text{Im}G_R(\omega)$ \eqref{deno} we have desired relation 
\begin{align}\label{eq:BM:21}
\xi^0(\omega )=f_{\omega}(r_h)\xi^h(\omega)
\end{align} 
For the AdS$_3$-BH system we are considering we can exactly calculate $f_{\omega}(r_h)$ from \eqref{eq:TL:mode} and equation \eqref{eq:BM:21} reduces to
\begin{align}\label{eq:BM:22}
\xi^0(\omega )= \left(1-i\frac{\omega}{\pi T}\right) \xi^h(\omega) 
\end{align} 

Once we have obtained this relation \eqref{eq:BM:21} we can use the horizon fluctuation-dissipation theorem\eqref{eq:dampedFD} and \eqref{eq:BM:19} to get
\begin{align}\label{eq:BM:23}
\langle\xi^0(-\omega)\xi^0(\omega) \rangle = -(1+2n_B)\text{Im}G_R(\omega)
\end{align} 
This is the statement of boundary fluctuation-dissipation theorem. \\

\section{Different Time Scales} 
\label{sec:Time scales}

Brownian motion is usually characterized mainly by two time scales \cite{Kubo2} : relaxation time(t$_{\text{r}}$) and collision time(t$_{\text{c}}$). Apart from these two there is another time scale called mean free path time(t$_{\text{mfp}}$). These different time scales come very naturally in kinetic theory of fluids. 
 But in the  holographic context when one considers classical gravity in the bulk the dual field theory is inevitably strongly coupled. In the same spirit the fluid that contains the quark in present context is strongly coupled. As a consequence, as we will see in this section, all those intuitive notions from kinetic theory don't go through in this case of holographic Brownian motion. First of all we will define the different time scales mentioned above. \\

\begin{itemize}

\item The \emph{relaxation time} is a time scale which separates ballistic regime where the Brownian particle moves inertially (displacement $\sim$ time) from diffusive regime where it undergoes a random walk (displacement $\sim$ $\sqrt{\text{time}}$). This is the time taken by the system to thermalize. 
In the small frequency regime \eqref{eq:TL:GF5} (using \eqref{eq:Mass}) we can write the Langevin equation \eqref{eq:LE:17} as 
\begin{align}\label{relax0}
\left[- \frac{\mu}{2 \pi} \omega^2 -i\gamma ~\omega + \ldots  \right] x(\omega)= ~\xi(\omega)
\end{align}
One obtains usual ballistic motion when the inertial term dominates over the diffusive term i.e,  $\frac{\mu}{2 \pi} \omega^2 >>\gamma ~\omega $. Evidently one gets one characteristic frequency when these two terms are of equal strength. 
\begin{align*}
\omega \sim \frac{\gamma}{\mu} \approx \frac{\sqrt{\lambda}T^2}{\text{M}_{\text{kin}}}
\end{align*}
Consequently the corresponding time scale (relaxation time) 
\begin{align}\label{relax}
\text{t}_\text{r} \sim \frac{\text{M}_{\text{kin}}}{ \sqrt{\lambda}T^2}
\end{align}

\item The \emph{collision time} is defined as a time scale over which random noise is correlated or in other words it's the time elapsed in a single collision. It measures how much $\gamma(t-t')$ deviates from $\delta(t-t')$. From \eqref{eq:TL:GF3.1} we obtain
\begin{align}\label{memory1}
\gamma(t)\equiv G_R(t)= \frac{\mu^2}{\sqrt{\lambda}}\left(-\frac{\mu^2}{\lambda}+4 \pi^2 T^2\right) e^{-\frac{\mu}{\sqrt{\lambda}} t}
\end{align} 
 In \eqref{memory1} $\frac{\sqrt{\lambda}}{\mu}$ naturally comes out to be a time scale. This is the `memory time' which  fixes the width of $\gamma(t)$. Hence it determines the duration of collision (t$_{\text{c}}$).\footnote{One can observe that this is the pole of the retarded Green function at $\omega = -i \frac{\mu}{\sqrt{\lambda}} $ that fixes the collision time scale. Here we have used Dirichlet boundary condition (standard quantization) \eqref{eq:TL:mode} on the string modes at the boundary. On the other hand, if one uses Neumann boundary condition (alternative quantization), presumably one would get the retarded Green function, $\tilde{G}_R =  G_R^{-1}$ (see Appendix A of \cite{FLR}). These two different boundary conditions describe two completely different dual CFTs. Therefore, for the latter CFT the corresponding time scale is determined by the \emph{zero} of our Green function G$_R(\omega)$ i.e, $\omega = -i 2\pi T$ and thus $\text{t}_\text{c} \sim \frac{1}{T}$. For example, in \cite{Mukund} Neumann boundary condition is used and the collision time, $\text{t}_\text{c} \sim \frac{1}{T}$. }
\begin{align}\label{memory2}
\text{t}_\text{c}  = \frac{\sqrt{\lambda}}{\mu} \approx \frac{\sqrt{\lambda}}{\text{M}_{\text{kin}}}
\end{align} 

From \eqref{relax} and \eqref{memory2} 
\begin{align}\label{memory3}
\frac{\text{t}_\text{r}}{\text{t}_\text{c}} \sim \left(\frac{\text{M}_{\text{kin}}}{\sqrt{\lambda}T}\right)^2 
\end{align} 

For `dilute' fluid one expects  t$_{\text{r}} >> \text{t}_\text{c}$. But from \eqref{memory3} it is clear that for strongly coupled fluids for which $\lambda >> 1$ this relation is not necessarily true.\\

\item The \emph{mean free path time} is the time elapsed between two consecutive collisions of the Brownian particle. As argued in \cite{Mukund} to obtain t$_{\text{mfp}}$ one needs to compute the 4-point correlation function. Therefore it will be suppressed by a factor of $\frac{1}{\sqrt{\lambda}}$ compared to $\text{t}_\text{c}$. 

\begin{align}\label{mfp}
\text{t}_\text{mfp} \sim  \frac{1}{\sqrt{\lambda}}\text{t}_\text{c} \approx \frac{1}{\text{M}_{\text{kin}}}
\end{align}

Again from \eqref{mfp} and \eqref{memory2} $\text{t}_\text{mfp} >> \text{t}_\text{c}$ does not necessarily hold for $\lambda>>1$. 
\end{itemize}


\section{Conclusions} 
\label{sec:Conclusions}

We have studied the Brownian diffusion of a particle in one dimension using the holographic techniques. The holographic dual is a BTZ black hole with a string.  We have used the Green  function techniques of Son and Teaney \cite{Son-Teaney}.   Since the differential equation can be solved exactly  we find an exact Green function and an exact (generalized) Langevin equation. \\

Some interesting features:\\
\begin{itemize}
\item 
We show that the exact generalized Langevin equation, which is valid on short time scales also, does not suffer from the inconsistency that is associated with the usual Langevin equation that has a delta function for the drag term. \\

\item
We also find that the temperature dependent mass correction is zero (in the limit that the UV cutoff is taken to infinity) unlike in the higher dimensional cases. \\

\item
There is also a temperature independent dissipation at all frequencies. At high frequencies it is a drag term. This does not violate Lorentz invariance as the force on a quark moving with constant velocity for all time continues to be zero. This has already been studied in higher dimensional systems and is due to radiation \cite{Mikhailov,Chernicoff1,Chernicoff2,Gubser2,CT2,Chernicoff}. It is noteworthy that a temperature independent dissipation in one dimension has also been noted in the condensed matter literature \cite{Pitaevskii,Mohanty,Bonart,Cherny,Walmsley,Kamenev}.\\

\item
Once again because an exact Green function is available, the ``stretched horizon'' can in fact be placed at an arbitrary radius and an effective action obtained. \\

\end{itemize}

It would be interesting to study the holographic RG interpretation \cite{Strominger,FLR} in this case. It would also be interesting to study the same problem using a charged BTZ black hole, thereby introducing a chemical potential. The phenomenon of zero temperature dissipation and drag is fascinating and it would be interesting to explore this further using holographic techniques.

\section*{Acknowledgments}

It is a pleasure to thank G. Baskaran and S. Kalyana Rama for very useful discussions. P.B. acknowledges a helpful discussion with Ronojoy Adhikari. \\ 

\paragraph{Note added.} After this work was completed we came across a paper \cite{Atmaja} that deals with very
similar issues.


\appendix

\section{Associated Legendre Differential Equation and Its Solutions}
\label{sec:A}

The associated Legendre differential equation is a generalization of the Legendre differential equation and is given by
\begin{align}\label{eq:A:DE1}
\frac{d}{dz}\left[(1-z^2)\frac{dy}{dz}\right]+\left[\lambda(\lambda + 1)+\frac{\mu^2}{1-z^2}\right]y=0
\end{align}
which can be written 
\begin{align}\label{eq:A:DE2}
(1-z^2)y^{\prime \prime}-2zy^{\prime}+\left[\lambda(\lambda + 1) - \frac{\mu^2}{1-z^2}\right]y=0
\end{align}
$P^\mu_\lambda$ and $Q^\mu_\lambda$ are the two linearly independent solutions to the associated Legendre D.E . The solutions $P^\mu_\lambda$ to this equation are called the associated Legendre polynomials (if $\lambda$ is an integer), or associated Legendre functions of the first kind (if  is not an integer). Similarly, $Q^\mu_\lambda$ is a Legendre function of the second kind. 
These functions may actually be defined for general complex parameters and argument. In particular they can be expressed in terms of hypergeometric functions and gamma functions
\begin{align}\label{eq:A:P}
P^\mu_\lambda(z)&=\frac{1}{\Gamma (1-\mu)}\left[\frac{1+z}{1-z}\right]^{\mu/2} {}_2F_1\left(-\lambda,\lambda+1;1-\mu;\frac{1-z}{2}\right) \\ \nonumber \\\label{eq:A:Q}
Q^\mu_\lambda(z)&=\frac{\sqrt{\pi}~\Gamma (\lambda+\mu+1)}{2^{\lambda+1}\Gamma (\lambda+3/2)}\frac{1}{z^{\lambda+\mu+1}}\left(1-z^2\right)^{\frac{\mu}{2}}{}_2F_1\left(\frac{\lambda+\mu+1}{2},\frac{\lambda+\mu+2}{2};\lambda+\frac{3}{2};\frac{1}{z^2}\right)
\end{align}

For the EOM of the string \eqref{eq:TS:EOM} the general solution will be 
\begin{align}\label{eq:A:Soln}
f_\omega(r)= C_1 \frac{P^{i\mathfrak{w}}_1}{r} + C_2 \frac{Q^{i\mathfrak{w}}_1}{r}
\end{align}
From above expressions \eqref{eq:A:P} and \eqref{eq:A:Q} we get, 
\begin{align*}
P^{i\mathfrak{w}}_1(r)&=\frac{1}{\Gamma (1-i\mathfrak{w})}\left[\frac{1+r}{1-r}\right]^{i\mathfrak{w}/2} {}_2F_1\left(-1,2;1-i\mathfrak{w};\frac{1-r}{2}\right) \\ \\
Q^{i\mathfrak{w}}_1(r)&=\frac{\sqrt{\pi}~\Gamma (2+i\mathfrak{w})}{4\Gamma (5/2)}\frac{1}{r^{(2+i\mathfrak{w})}}\left(1-r^2\right)^{\frac{i\mathfrak{w}}{2}}{}_2F_1\left(\frac{2+i\mathfrak{w}}{2},\frac{3+i\mathfrak{w}}{2};\frac{5}{2};\frac{1}{r^2}\right)
\end{align*}

Near the horizon $f(r)=1-\frac{1}{r^2}\to 0$ i.e, $r \to 1$. So the dominant behavior of the solution near the horizon will be of the form, $$f_\omega(r)\sim (r-1)^{\pm\alpha}$$ From above two solutions it is evident that $\frac{P^{i\mathfrak{w}}_1(r)}{r}\sim (1-r)^{-i\mathfrak{w}/2} $ and $ \frac{Q^{i\mathfrak{w}}_1(r)}{r}\sim (1-r)^{i\mathfrak{w}/2} $. 
Now, 
\begin{align*}
&e^{-i\omega t}.(1-r)^{-i\mathfrak{w}/2} \sim e^{-i\omega\{t+\frac{1}{2\pi T}ln(r-1)\}}
\end{align*}
Notice that near the horizon $ r \to 1$ ,  $ln(r-1)$ goes more and more negative. Now to keep the phase fixed $t$ must increase. That means this wave moves towards the horizon with increment of time . So, $\frac{P^{i\mathfrak{w}}_1}{r} \sim (1-r)^{-i\mathfrak{w}/2}$ is the desired \textit{incoming wave} solution . And by the same token $\frac{Q^{i\mathfrak{w}}_1}{r} \sim (1-r)^{i\mathfrak{w}/2}$ is the \textit{outgoing wave} solution. To get retarded propagator one has to choose $\frac{P^{i\mathfrak{w}}_1}{r}$.

\section{Retarded Bulk to Bulk Correlators}
\label{sec:B}

If we have retarded Green function in the boundary theory we can always construct other Green functions. As we know the solution to the string EOM exactly we can build the exact retarded bulk-to-bulk correlator, $G_{\text{ret}}(\omega,r, \tilde{r})$ which is in-falling at the horizon and normalizable at the boundary. In section \ref{sec:TS} we had $f_\omega(r)$ as a solution to the wave equation \eqref{EOM} and so was $f^*_\omega(r)$. As it was a linear differential equation any linear combination of them e.g, Im$f_\omega(r)=\frac{f_\omega(r)-f^*_\omega(r)}{2i}$ is also a solution. But we had chosen them such that $f_\omega(r)$ and $f^*_\omega(r) \to 1 $ as $r \to \infty $ . Therefore Im$f_\omega(r)$ is a normalizable solution to that wave equation \eqref{EOM}. Thus the retarded bulk-to-bulk correlator is defined as 
\begin{align} \label{G_ret}
G_{\text{ret}}(\omega,r, \tilde{r})=\frac{\text{Im} f_\omega(r)f_\omega(\tilde{r})\theta(r,\tilde{r})+f_\omega(r)\text{Im}f_\omega(\tilde{r})\theta(\tilde{r},r)}{T_0(\tilde{r}) W_\text{ret}(\tilde{r})}
\end{align}
  
\begin{align} 
W_\text{ret}(\tilde{r})\equiv \text{Im}f^{\prime}_\omega(\tilde{r}) f_\omega(\tilde{r})-f^{\prime}_\omega(\tilde{r})\text{Im}f_\omega(\tilde{r})
\end{align}
Now from equation \eqref{eq:TL:mode} (taking $r_m \to \infty$) we obtain the Wronskian as 

\begin{align} \label{Wronskian}
W_\text{ret}(\tilde{r})= \frac{ \mathfrak{w}^3+\mathfrak{w} }{\tilde{r}^2-\tilde{r}^4} 
\end{align}

The interesting thing to notice that the Wronskian depends on $\tilde{r}$ in such a way that $T_0(\tilde{r})= \frac{\sqrt{\lambda}\pi^2 T^3}{2}~\tilde{r}^2(\tilde{r}^2-1)$ cancels that $\tilde{r}$-dependence. Therefore the denominator of \eqref{G_ret} 
\begin{align} \label{deno}
T_0(\tilde{r})W_\text{ret}(\tilde{r})= ~-\frac{1}{2} \pi ^2 \sqrt{\lambda } T^3 \left(\mathfrak{w} ^3+\mathfrak{w} \right)= ~+\text{Im}G_R(\omega)
\end{align}
becomes independent of $\tilde{r}$. \\

Now we can write \eqref{G_ret} as 
\begin{align}
G_{\text{ret}}(\omega,r, \tilde{r}) \equiv~ G^+_{\text{ret}}(\omega,r, \tilde{r})~\theta(r,\tilde{r})~+~G^-_{\text{ret}}(\omega,r, \tilde{r})~\theta(\tilde{r},r)
\end{align}

And finally using \eqref{eq:TL:mode} and \eqref{deno} we obtain  
\begin{align}
G^+_{\text{ret}}(\omega,r, \tilde{r})= \left(1-r^2\right)^{-\frac{i \omega }{2 \pi  T}}&\left(\frac{1+\tilde{r}}{1-\tilde{r}}\right)^{\frac{i \omega }{2 \pi  T}} (\pi \tilde{r}  T-i \omega )~e^{-\frac{\omega }{T}}  \nonumber\\ 
&\frac{ (\omega +i \pi  r T) ~(1+r)^{\frac{i \omega }{\pi  T}}+e^{\omega /T} ~(1-r)^{\frac{i \omega }{\pi  T}} ~(\omega -i \pi  r T) }{\pi  \sqrt{\lambda } r \tilde{r} T^2 \omega ~ \left(\pi ^2 T^2+\omega ^2\right)} 
\end{align}
\begin{align}
G^-_{\text{ret}}(\omega,r, \tilde{r})= \left(1-\tilde{r}^2\right)^{-\frac{i \omega }{2 \pi  T}}&\left(\frac{1+r}{1-r}\right)^{\frac{i \omega }{2 \pi  T}} (\pi r  T-i \omega )~e^{-\frac{\omega }{T}}  \nonumber\\ 
&\frac{ (\omega +i \pi  \tilde{r} T)~ (1+\tilde{r})^{\frac{i \omega }{\pi  T}}+e^{\omega /T} ~(1-\tilde{r})^{\frac{i \omega }{\pi  T}}~ (\omega -i \pi  \tilde{r} T) }{\pi  \sqrt{\lambda } r \tilde{r} T^2 \omega  ~\left(\pi ^2 T^2+\omega ^2\right)} 
\end{align}


\vspace{2cm}
\centerline{***}
\end{document}